\documentclass[prb,aps,twocolumn,nofootinbib,superscriptaddress]{revtex4-2}
\usepackage[usenames,dvipsnames]{color}
\usepackage[colorlinks=true,linkcolor=Blue,citecolor=Blue,urlcolor=Blue]{hyperref}
\usepackage{comment,graphicx,physics,mathrsfs}
\usepackage{bbm}
\usepackage{bm}
\usepackage{amsmath}
\usepackage{amssymb}
\usepackage{mathptmx}
\usepackage[version=3]{mhchem}
\usepackage{soul}

\newcommand{\pa}[1]{\left( #1 \right)}

\definecolor{orange}{rgb}{1,0.37,0.12}

\renewcommand\vec{\mathbf}

\newcommand{\agme}{\alpha^{\text{GME}}}

\begin{document}
\title{Gyrotropic magnetic effect in metallic chiral magnets}
\author{Nisarga Paul}
\affiliation{Department of Physics, Massachusetts Institute of Technology, Cambridge, Massachusetts 02139, USA}
\affiliation{Kavli Institute for Theoretical Physics, University of
  California, Santa Barbara, CA 93106, USA}
\author{Takamori Park}
\affiliation{Department of Physics, University of California, Santa Barbara, CA 93106, USA}
\author{Jung Hoon Han}
\affiliation{Department of Physics, Sungkyunkwan University, Suwon 16419, South Korea}
\author{Leon Balents}
\affiliation{Kavli Institute for Theoretical Physics, University of
  California, Santa Barbara, CA 93106, USA}
\affiliation{French American Center for Theoretical Science, CNRS,
  KITP, Santa Barbara, CA 93106-4030, USA}
\affiliation{Canadian Institute for Advanced Research, Toronto, Ontario, M5G 1M1, Canada} 

\begin{abstract}
    We study the gyrotropic magnetic effect (GME), the low-frequency limit of optical gyrotropy, in metals and semimetals coupled to chiral spin textures. In these systems, the chiral spin texture which lacks inversion symmetry can imprint itself upon the electronic structure through Hund's coupling, leading to novel low-frequency optical activity. Using perturbation theory and numerical diagonalization of both relativistic and non-relativistic models of conduction electrons coupled to spin textures, we analyze how the GME manifests in both single-$q$ and multi-$q$ textures. Analytical expressions for the rotatory power are derived in terms of universal scaling functions. Estimates based on realistic material parameters reveal an experimentally viable range of values for the rotatory power. The GME arises from the orbital and spin magnetic moments of conduction electrons, with the orbital part closely tied to Berry curvature and playing a significant role in relativistic metals but not so in non-relativistic metals where there is no inherent Berry curvature. The spin contribution to the GME can be significant in non-relativistic metals with a large Fermi energy. Our work shows that the GME can be a sensitive probe of magnetic chirality and symmetry breaking in metallic chiral magnets.  
\end{abstract}
\maketitle

\textit{Introduction.}--- Optical gyrotropy refers to the rotation of the axis of polarization of light passing through a medium lacking inversion $(\mathcal{P})$ symmetry~\cite{Landau1984,Agranovich,Jerphagnon1976Aug,Cheong2019Oct}. First discovered in quartz by Arago in 1811 \cite{arago1811}, it has found widespread application in studies of RNA, DNA, sugars, and chiral nematic liquid crystals. In condensed matter systems, gyrotropic effects are among the most sensitive and definitive probes of symmetry breaking, making them invaluable in identifying phase transitions and characterizing symmetry-broken phases of matter ~\cite{Landau1984,Agranovich,Jerphagnon1976Aug}. A related phenomenon is the Faraday/Kerr rotation of light polarization upon transmission/reflection off a medium which breaks time-reversal $(\mathcal{T})$ symmetry with nonzero Hall conductivity~\cite{fried14,MnGe-NatComm21}. On the other hand, the low-frequency limit ($\omega \rightarrow 0$) of optical gyrotropy, known as the gyrotropic magnetic effect (GME), can develop in a medium with only broken $\mathcal{P}$. The rotatory power--- the angle of rotation of light per unit length--- vanishes as $\omega^2$ in insulators but can attain a constant value in materials with gapless excitations.
\par 

Optical gyrotropy in chiral molecules and crystal insulators is a classic phenomenon with solid theoretical understanding~\cite{Landau1984,Agranovich,barron09,Cheong2019Oct}. On the other hand, the theory of GME in chiral metals has been established relatively recently~\cite{pesin15,Goswami2015Oct,souza16}, with a pure GME in the absence of an applied current not yet confirmed experimentally~\cite{shalygin2012current,Tsirkin2018Jan}. Given the potential of the GME as a highly sensitive probe of crystal symmetry by optical means, it seems only fitting to broaden the search for material candidates for its observation.

Chiral magnets are a broad class of materials that, besides breaking $\mathcal{T}$ as a result of magnetic order, typically break $\mathcal{P}$ owing to noncollinear spin textures that lack an inversion center~\cite{Cheong2022Apr}. 
They have been extensively studied due to their ability to host a range of topological solitons with potential spintronics applications, from skyrmions and merons in two dimensions to hedgehogs in three dimensions, which can be easily accessed by variations in temperature and applied magnetic field~\cite{Nagaosa2013Dec,Fert2017,Han2017,Yang2021May,Tokura2020Nov}. Despite these advances in identifying and controlling magnetic phases of chiral magnets, the investigation of their optical gyrotropy~\cite{Cheong2019Oct} as a direct manifestation of the magnetic chirality remains quite limited thus far~\cite{Masuda2021Apr}. 
\par 
\begin{figure}
    \centering  \includegraphics[width=0.8\linewidth]{figure_schematic_v4.png}
    \caption{Schematic of the gyrotropic magnetic effect (GME) in chiral magnets. Linearly polarized light (blue waves) enters a material whose spin texture breaks inversion symmetry. The conduction electrons couple to the local spins via a Hund’s exchange term and induce rotation of the light’s polarization upon transmission. Measuring the GME could provide a sensitive probe of the chiral magnetic order.}
    \label{fig:schematic}
\end{figure}
We propose metallic chiral magnets, i.e. spin-textured metals and semimetals that lack inversion symmetry in their magnetic structure, as the new potential platform for observing the GME (Fig.~\ref{fig:schematic}). Early theoretical investigations of the GME focused on non-magnetic Weyl semimetals (WSMs)~\cite{pesin15,Goswami2015Oct,souza16,Tsirkin2018Jan,grushin18,Wang2019Dec,Wang2020May} in connection with the sought-after chiral magnetic effect~\cite{son-yamamoto,burkov12,son-spivak,niu13}. We show, in fact, a relativistic electron dispersion is not necessary for the observation of the GME, and it also occurs in non-relativistically dispersing electrons coupled to a chiral spin texture, greatly broadening the class of material candidates. Instead of $\mathcal{P}$-breaking induced by structural asymmetry~\cite{Goswami2015Oct,pesin15,souza16}, we consider situations where such asymmetry arises from the spin texture and affects the electronic structure through Hund's coupling of local and itinerant spins. We propose that the GME exists as a generic feature in all such metallic chiral magnets and could provide a sensitive diagnostic for detecting and characterizing the complex magnetic orders therein. 
\\

\textit{Model.}--- A minimal model describing electrons coupled to a spin texture is given by
\begin{equation}\label{eq:H}
H = H_0(\vec p) + J\sum_{\vec r} \bm S_{\vec r} \cdot \bm\sigma 
\end{equation}
where $H_0$ captures the electronic dispersion, $J>0$ is the Hund's coupling to a static local spin texture $\bm S_{\vec r}$, and $\bm \sigma=(\sigma^x,\sigma^y,\sigma^z)$ are the Pauli operators for itinerant electron spins. This type of model has served as the canonical means of studying the properties of chiral magnets coupled to itinerant electrons in the past~\cite{Nagaosa2013Dec,Han2017}. We will consider both non-relativistic and relativistic dispersions: 
\begin{equation}
    H_0^{\text{NR}} = {\bf p}^2/2m,\qquad H_0^{\text{R}} = \sum_{w} \chi_w  v_F (\vec p \cdot \bm \sigma )_w . 
    \label{eq:H0}
\end{equation}
For $H_0^{\text{NR}}$, $m$ is an effective mass and ${\bf p}$ is measured from the band bottom. For $H_0^{\text{R}}$, $v_F$ is the Fermi velocity at the Weyl points, $\sum_w$ is a direct sum over Weyl points with chiralities $\chi_w$, and ${\bf p}$ is measured from the location of respective Weyl point. For simplicity we will restrict our analysis to a pair of $\mathcal{P}$-conjugate Weyl points, located at ${\bf k}_0$ and ${\mathcal P}{\bf k}_0 = -{\bf k}_0$ with respective chiralities $\chi_w=\pm 1$. 
 Moreover, we will assume that the spin texture is periodic: $    \bm S_{\vec r} =\sum_{\vec G} \bm S_{\vec G} e^{i \vec G \cdot \vec r}$ where $\vec G$ belongs to a reciprocal lattice $\Lambda$, allowing us to work in the momentum space Brillouin zone associated to $\Lambda$. 

\textit{Optical rotation.}--- When linearly polarized light passes through an optically active medium, 
 its polarization is rotated by an angle $\phi$ proportional to the thickness $d$. The angle $\phi$ depends on the GME tensor, which at low frequencies $\hbar\omega\ll\varepsilon_{\text{gap}}$ (where $\varepsilon_{\text{gap}}$ is the gap to remote bands) takes the form 
\begin{equation}\label{eq:alphaGME}
    \agme_{ij} = \frac{i\omega\tau e}{1-i\omega \tau} \sum_{n} \int [d\vec k] \frac{\partial f}{\partial \varepsilon_{\vec k n}} v_{\vec kn,i}m_{\vec kn,j} ,
\end{equation}
capturing intraband contributions for a metal with relaxation time $\tau$. The integral is over the Brillouin zone with $[d\vec k] = d^3\vec k/(2\pi)^3$, $f=f(\varepsilon_{n\vec k})$ is the distribution function, and $\varepsilon_{\vec k n}, \vec v_{\vec kn} = \partial \varepsilon_{\vec k n} /\hbar\partial \vec k$, and $\vec m_{\vec kn}$ are band energies, velocities, and magnetic moments, respectively. The magnetic moments are $\vec m_{\vec kn} = \vec m_{\vec k n}^{\text{spin}} + \vec m_{\vec k n}^{\text{orb}}$ with
\begin{subequations}
\begin{align}
        \label{eq:mkn}
        \vec m_{\vec k n}^{\text{spin}}&= -\frac{eg_s}{2m_e} \langle u_{\vec kn}|\frac{\hbar\bm\sigma}{2} |u_{\vec k n}\rangle\\
        \vec m_{\vec k n}^{\text{orb}}  &= -i\frac{e}{2\hbar}\langle \bm \nabla_{\vec k}u_{\vec kn}|\times (H_{\vec k} - \varepsilon_{\vec kn})|\bm \nabla_{\vec k} u_{\vec k n}\rangle 
\end{align}
\end{subequations}
\noindent 
for Bloch states $| u_{\vec k n} \rangle$, where $g_s \approx 2$ is the electron $g$-factor and $m_e$ is the bare electron mass. The {\it rotatory power} is then
\begin{equation}
    \rho = \frac{\phi}{d} = \frac{1}{2c^2\varepsilon_0}\Re\pa{ \agme_{ij}\hat n_i \hat n_j - \bar\alpha^{\text{GME}}}
            \label{optical-activity-formula} 
\end{equation}
where $\bar \alpha^{\text{GME}} = \Tr\agme$, for light propagating along the unit wave normal $\hat{\mathbf{n}}$. Here $c$ is the speed of light and $\varepsilon_0$ is the vacuum permittivity. As an example, $\rho \sim 0.328$ rad/mm in quartz for light of wavelength $\lambda = 0.63\mu$m \cite{Newnham2005}. The rotatory power vanishes as $\omega^2$ in insulators, molecules, and dirty metals, while in clean metals ($\omega \tau \gg 1$) it goes to a constant at low frequencies~\cite{supp}. 

\par 

\textit{Results.}--- When the Hund's coupling $J$ is weak compared to the characteristic kinetic energy of the electrons, the GME tensor can be calculated with perturbation theory as outlined in \cite{supp}. We separate the orbital and spin contributions to the GME tensor, writing
\begin{equation}\alpha_{ij}^{\textrm{GME,D}}=\alpha_{ij}^{\textrm{GME,D,orb}}+\alpha_{ij}^{\textrm{GME,D,spin}}
\end{equation}
for either dispersion, D = R or NR. The leading contribution to the GME tensor occurs at $O(J^2)$ and takes the form 
\begin{widetext}


\begin{subequations}
\begin{align}
\alpha_{ij}^{\textrm{GME,NR,spin}} &= \frac{i\omega \tau}{1-i\omega\tau}  \alpha_0 \frac{g_s J}{4 m_e (v_F^{\rm NR} )^2} \sum_{\vec G\neq 0} \,  f_{\vec G , ij} g^{\text{NR}}(\xi) \,\,    +\cdots\quad  \label{eq:nrperturbspin}  \\
\alpha_{ij}^{\textrm{GME,R,orb}}&=
  \frac{i\omega \tau}{1-i\omega\tau}  \alpha_0 \sum_{\vec G\neq0}\frac{J}{\hbar v_F G} 
  f_{\vec G} \qty[g_{\perp}^\textrm{R,orb}\qty(\xi) P^\perp_{\vec G , ij}
+g_{\|}^\textrm{R,orb}\qty(\xi)  P^\parallel_{\vec G , ij} ] +\cdots \label{eq:perturborb} \\
\alpha_{ij}^{\textrm{GME,R,spin}} &= \frac{i\omega \tau}{1-i\omega\tau}  \alpha_0 \frac{g_s J}{4 m_e v_F^2}
  \sum_{\vec G\neq0} \bigg( f_{\vec G}\qty[g_{\perp}^\textrm{R,spin}\qty(\xi) P^\perp_{\vec G , ij}
+g_{\|}^\textrm{R,spin}\qty(\xi) P^\parallel_{\vec G , ij} ] +  f'_{\vec G , ij} g_L\qty(\xi) \bigg)  +\cdots.  \label{eq:perturbspin}
\end{align}
\label{eq:perturb}
\end{subequations}
\end{widetext}
The prefactor $\alpha_0  = \frac{e^2}{h^2}J$ carries the dimension of the GME tensor. We introduced the non-relativistic velocity $v_F^{\rm NR} = \hbar k_F /m$ in terms of the Fermi momentum $k_F$ (Fermi energy $E_F = \hbar^2 k_F^2 /2m$) and the renormalized electron mass $m$. The Fermi momentum $k_F$ is related to the Fermi energy $E_F$ in the relativistic case as $E_F = \hbar v_F k_F$. In deriving the above formulas, the dimensionless quantity $\xi = 2k_F/G$ ($G=|{\bf G}|$) is assumed to satisfy $|\xi|<1$ to rule out scattering among the same-energy states and the opening of the minigap. The assumption enables us to focus on the lowest-energy bands in the non-relativistic case or the bands closest to the Weyl node as being active. Terms in $\cdots$ are subleading in $J/E_K^{\text{NR}}, J/E_K^{\text{R}},$ and $J/m_ev_F^2$, where $E_K^{\text{NR}} = \hbar^2 G^2/2m$ or $E_K^{\text{R}} = \hbar v_F G$ is the characteristic electronic kinetic energy. We refer to these quantities as $E_K$ when the context is clear. 

Form factors $f_{\vec G , ij } = i \vec{\hat{G}}_i (\bm S_{\vec G} \times \bm S_{-\vec G})_j $ and $f_{\bf G} = \Tr [f_{\vec G,ij} ]$ are introduced in Eq.~\eqref{eq:perturb} in terms of $\vec{\hat{G}}=\vec{G}/G$, where $G = 2\pi /\lambda$ is related to the spin texture wavelength $\lambda$. Transverse and longitudinal projectors are defined by $P^\perp_{\vec G , ij} = \delta_{ij} - \vec{\hat{G}}_i\vec{\hat{G}}_j$ and $P^\parallel_{\vec G, ij} = \vec{\hat{G}}_i \vec{\hat{G}}_j$. The form factor $f'_{\vec G , ij} \equiv f_{\vec G , ij} - f_{\vec G} P^\parallel_{\vec G , ij}$ appearing in the last equation is equivalent to $i \vec{\hat G}_i \qty[{\bm S}_{\vec G} \times {\bm S}_{-\vec G} ]^\perp_j$ where $\qty[ {\bm S}_{\vec G} \times {\bm S}_{-\vec G} ]^\perp = {\bm S}_{\vec G} \times {\bm S}_{-\vec G} -  \vec{\hat G}\cdot \qty( {\bm S}_{\vec G} \times {\bm S}_{-\vec G} ) \vec{\hat G}$ refers to the transverse component. The form factor $f_{\vec G}$ shows up only in spin spirals rotating in the plane {\it perpendicular} to the propagation vector $\vec{G}$. Spin cycloids with spins rotating in the plane containing ${\bf G}$ may only contribute to $\alpha_{ij}^{\textrm{GME,R,spin}}$. The non-relativistic orbital magnetic moment and $\alpha_{ij}^{\text{GME,NR,orb}}$ are highly suppressed, presumably due to the absence of spin-orbit coupling and the lack of Berry curvature in the unperturbed bands. 

\begin{figure}
    \centering
\includegraphics[width=\linewidth]{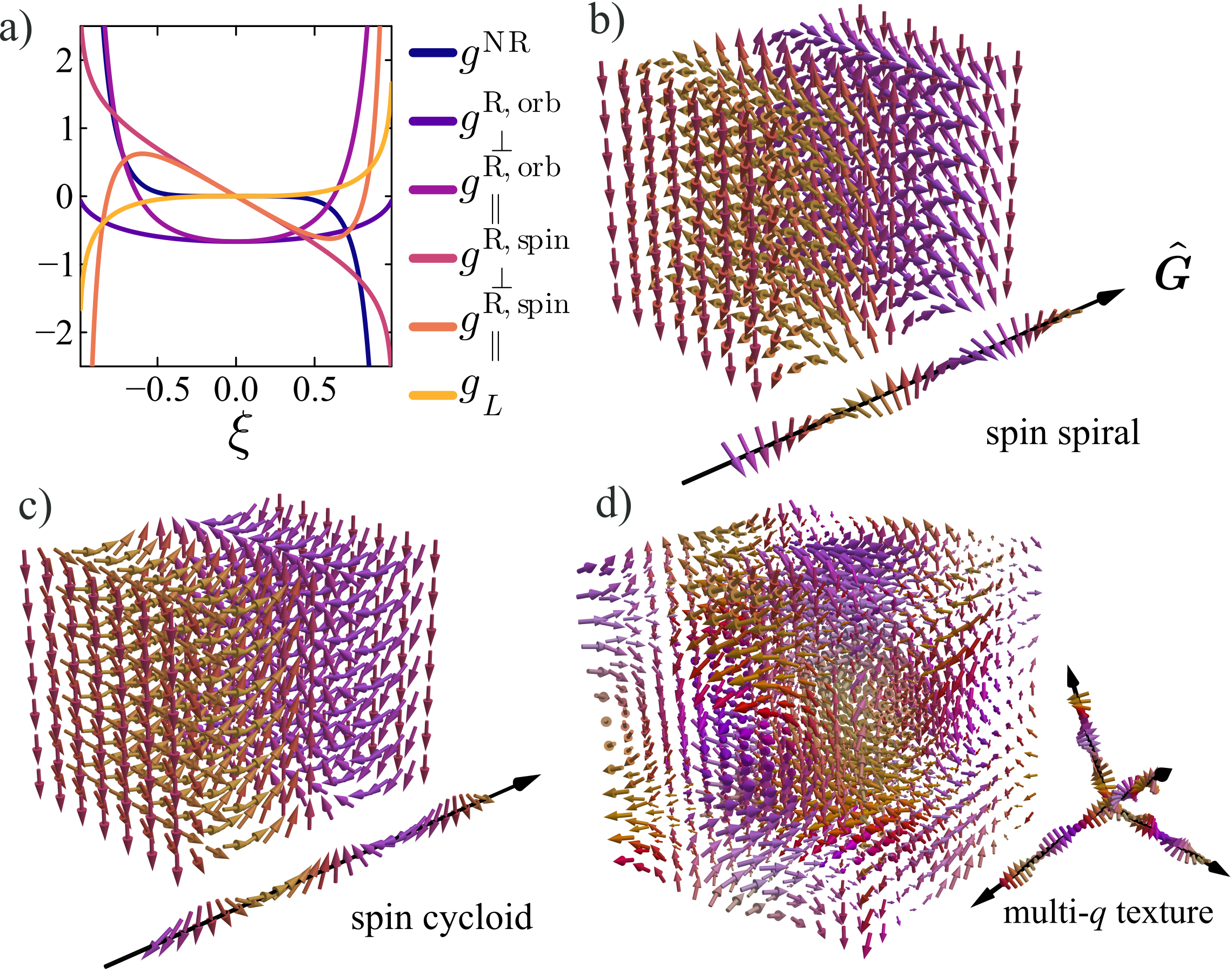}
    \caption{(a) Scaling functions occuring in Eq. \eqref{eq:perturb} for the GME tensor ($\xi = 2k_F/G$). (b-d) Three types of magnetic textures considered in this work: a spiral, cycloid, and multi-$q$ texture. Colored arrows represent the local magnetic moments $\bm S_{\vec r}$ and black arrows represent constituent wavevectors.}    \label{fig:scalingfns}
\end{figure}

A number of dimensionless scaling functions are introduced, which are given by
\begin{subequations}
\begin{align}
  g^{\text{NR}}(\xi) =& 2\xi^2 \left(\frac{\xi}{\xi^2-1} +\tanh^{-1}\xi \right)\\
  g_{\perp}^\textrm{R,orb}(\xi)=&2\qty(1-\xi^2)\qty(\xi-\tanh^{-1}\xi)/\xi^3\\
  g_{\|}^\textrm{R,orb}(\xi)=& 2\frac{\xi(\xi^4-4\xi^2+2)-(3\xi^4-5\xi^2+2)\tanh^{-1}\xi}{\xi^3(\xi^2-1)}\\
  g_{\perp}^\textrm{R,spin}(\xi)=&(\xi-\xi^3-\tanh^{-1}\xi)/\xi^2\\
  g_{\|}^\textrm{R,spin}(\xi)=&\frac{(2-3\xi^2)(\xi+(\xi^2-1)\tanh^{-1}\xi)}{\xi^2(\xi^2-1)}\\
  g_L(\xi)=&-\xi+\tanh^{-1}\xi
\end{align}
   \label{eq:scalingfns}
\end{subequations}
We plot them in Fig. \ref{fig:scalingfns}a. 
In general, the orbital part of the 
scaling functions are even in $\xi$ and finite at $\xi = 0$. The spin parts are odd in $\xi$ and vanish when $\xi = 0$. Only the $\xi \ge 0$ part is relevant for the non-relativistic scaling function $g^{\text{NR}}(\xi)$, whereas both signs of $\xi$ are meaningful for relativistic electrons since the Fermi energy can be of both signs. Spin contributions vanish for the Weyl electrons at charge neutrality ($E_F = 0$) while for non-relativistic electrons they remain finite since $\xi > 0$. The high-density region ($\xi > 1$) not captured by the scaling formulas can be treated by solving the Hamiltonian numerically. 

\begin{figure*}
    \centering
\includegraphics[width=\linewidth]{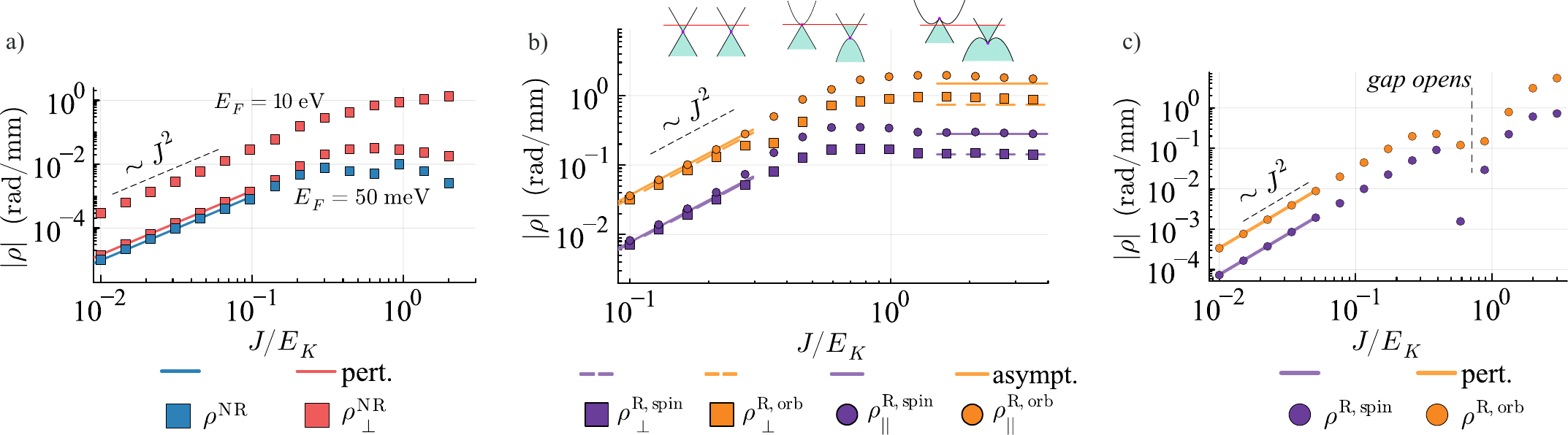}
\caption{(a) Rotatory power of non-relativistic metallic chiral magnet hosting a spin spiral (red) or face-centered cubic multi-$q$ spin texture (blue) with Hund's coupling $J$. Parameters used are $m = 0.1m_e, \lambda = 5$ nm ($E_K = h^2/2m\lambda^2 \approx$ 0.6 eV) for both high ($E_F=10$ eV, top curve) and low ($E_F=50$ meV, bottom curves) electron densities. 
(b) Rotatory power of a Weyl semimetal coupled to a spin spiral. Parameters used are $v_F = 10^{-3}c$, $\lambda = 5$ nm ($E_K = hv_F/\lambda\approx 0.25$ eV), and $E_F = 50$ meV. 
Upper insets heuristically depict bandstructures and Weyl nodes relative to $E_F$. (c) Rotatory power of a Weyl semimetal (with same parameters) coupled to a multi-$q$ spin texture.  
Numerical results (markers) are shown alongside analytical results from main text. }
    \label{fig:numerics}
\end{figure*}

 All scaling functions have comparable magnitudes as seen in Fig. \ref{fig:scalingfns}a, and one can make order-of-magnitude estimates of the dimensionless GME response $\alpha/ \alpha_0$ as
\begin{align} \frac{|\alpha|}{\alpha_0} \sim \frac{g_s J}{4 m_e (v_F^{\rm NR} )^2 } , ~~ \frac{J}{\hbar v_F G} , ~~ \frac{g_s J }{4 m_e v_F^2} 
\end{align}
for the three GME tensors in Eq.~\eqref{eq:perturb}, respectively. Assuming $v_F^{\rm NR} \sim v_F \sim 10^{-3} c$, the spin contributions to $| \alpha | / \alpha_0$ are $\sim J/(2 {\rm eV})$ for both electronic dispersions. The orbital contribution, unique to relativistic electrons, ranges between $J/(0.1 {\rm eV})$ and $J/(0.01 {\rm eV})$ for the spiral wavelength $\lambda \sim 10-100$ nm and dominates over the spin counterparts in the perturbative regime. It is also comparable to or larger than $\alpha = (e^2 /3h^2 ) \Delta \varepsilon$ obtained in \cite{souza16} for non-magnetic Weyl models where $\Delta \varepsilon$ is the Fermi energy difference between the two Weyl points, if we assume both $J\sim \Delta \varepsilon \sim 0.1$ eV. Contrary to \cite{souza16}, both Weyl nodes share the same Fermi level in our derivation of the GME response. 
Only a single pair of Weyl nodes is assumed in the derivation of the GME formulas. A second pair, related to the first by $\mathcal{T}$, would result in the same GME response and contribute additively. Our formulas for the relativistic electrons are applicable even in the case of Dirac electrons where a pair of Weyl nodes occur at the same momentum in the Brillouin zone. 
\par  


We illustrate our results using three representative examples of spin textures: a spin spiral (Bloch-like spiral), a spin cycloid (N\'eel-like spiral), and a multi-$q$ spiral texture. The \textit{spin spiral} has spins rotating in the plane orthogonal to the propagation vector $\vec G$ with chirality $\chi_s=\pm 1$:
\begin{equation}\label{eq:spinspiral}
    \bm S_{\vec r} = \bm S_{\vec 0} + \bm S_{\vec G} e^{i\vec G\cdot \vec r} + (\bm S_{\vec G})^* e^{-i\vec G\cdot \vec r} 
\end{equation}
where $\bm S_{\vec G} = (\hat{\vec{G}}_{\perp,1}+i\chi_s\hat{\vec G}_{\perp,2})/2$, $\hat{\vec G}_{\perp,1}\perp\hat{\vec G}_{\perp,2} $ and $\hat{\vec G}= \hat{\vec G}_{\perp,1}\times \hat{\vec G}_{\perp,2}$. The \textit{spin cycloid}, in contrast, rotates in a plane containing $\vec G$; it is given by Eq.~\eqref{eq:spinspiral} with $\bm S_{\vec G} = (\hat{\vec{G}}_{\perp,1}+i\chi_s\hat{\vec G})/2$. The multi-$q$ spiral texture, which forms a hedgehog or monopole-antimonopole lattice in real space, is a sum of spin spirals along various $\vec G$~\cite{Nagaosa2013Dec,MnGe-NanoLett15,Han2017}. As an explicit example we adopt a texture with face-centered cubic symmetry~\cite{supp}. Real-space configurations are shown in Fig.~ \ref{fig:scalingfns}. 

The rotatory power $\rho$ along high-symmetry directions of each spin texture can be worked out. For the spin spiral, using Eq.~\eqref{optical-activity-formula} combined with Eq.~\eqref{eq:perturb}, the rotatory power along the direction parallel to $\hat{\vec G}$ is $\rho_{\|}^{\text{NR}} = 0$ and
\begin{subequations}
    \begin{align}
        \rho_{\|}^{\text{R,orb}} &= \chi_s \rho_0 \frac{J}{\hbar v_FG}  g_{\perp}^{\text{R,orb}}(\xi)\\
        \rho_{\|}^{\text{R,spin}} &= \chi_s \rho_0 \frac{g_sJ}{4m_ev_F^2}  g_{\perp}^{\text{R,spin}}(\xi)
    \end{align}
    \label{eq:rhopar}
\end{subequations}
($\rho_0 = \alpha_0 / c^2\varepsilon_0$) while the rotatory power transverse to $\hat{\vec G}$ is 
\begin{subequations}
    \begin{align}
        \rho_{\perp}^{\text{NR}} &= \chi_s \frac{\rho_0}{2} \frac{g_sJ}{4m_e(v_F^{\text{NR}})^2}  g^{\text{NR}}(\xi)\label{eq:rhoNRperp}\\
        \rho_{\perp}^{\text{R,orb}} &= \chi_s \frac{\rho_0}{2} \frac{J}{\hbar v_FG}  [g_\perp^{\text{R,orb}}(\xi)+g_\|^{\text{R,orb}}(\xi)]\\
        \rho_{\perp}^{\text{R,spin}} &= \chi_s \frac{\rho_0}{2} \frac{g_sJ}{4m_ev_F^2} [ g_\perp^{\text{R,spin}}(\xi)+g_\|^{\text{R,spin}}(\xi)].
    \end{align}
    \label{eq:rhoperp}
\end{subequations}
In the non-relativistic case, only the transverse direction is optically active. For relativistic electrons, the orbital and spin contributions to the optical activity are of the same sign if $\xi >0$, indicating that electron-doped magnetic Weyl materials are likely to result in stronger rotatory power. The {\it sign} of the rotation angle depends on the spin chirality $\chi_s$ in all cases, implying that the optical rotation can be a sensitive diagnostic of the spin chirality in a metal. 
\par 

For the spin cycloid, the optical activity coefficient vanishes in both longitudinal and transverse directions due to parity symmetry. Symmetry-wise, this is due to the vanishing of the structure factor $i\hat{\vec G}\cdot (\bm S_{\vec G}\times \bm S_{-\vec G})$ for any spin texture with an inversion center. 

For the multi-$q$ texture, the high degree of symmetry results in an \textit{isotropic} rotatory power, taking the form
\begin{subequations}
    \begin{align}
    \rho^{\text{NR}} &= \chi_s \frac{\rho_0}{3} \frac{g_sJ}{4m_e(v_F^{\text{NR}})^2}  g^{\text{NR}}(\xi)\label{eq:rhoNR}\\   
    \rho^{\text{R,orb}} &= \chi_s \frac{\rho_0}{3} \frac{J}{\hbar v_f G}  \left[2g_\perp^{\text{R,orb}}(\xi) +g_{\|}^{\text{R,orb}}(\xi) \right]\\   
    \rho^{\text{R,spin}} &= \chi_s \frac{\rho_0}{3} \frac{g_sJ}{4m_ev_F^2}\left[2g_\perp^{\text{R,spin}}(\xi)+g_{\|}^{\text{R,spin}}(\xi)\right]
    \end{align}
    \label{eq:rhogen}
\end{subequations}
regardless of direction $\hat{\bf n}$. Again, electron-doped magnetic Weyl semimetals are likely to result in the largest rotatory power.
\par 

We plot the rotatory power for non-relativistic electrons coupled to the spin texture in Fig.~\ref{fig:numerics}a, using the perturbative formulas of the preceding discussion and numerical calculations. An excellent fit to the theoretical curve is found at small $J/E_K$ and small Fermi energy $E_F$. Outside this perturbative regime we rely on the numerical approach and find values comparable to that of quartz for $E_F \sim 10$ eV and $J \sim 1$ eV, typical of Hund-coupled metals. Further details of the numerical approach can be found in \cite{supp}. \par 

The rotatory power for Weyl electrons coupled to a spin spiral is plotted in Fig.~\ref{fig:numerics}b. Increasing as $J^2$ at small $J/E_K$, it eventually saturates to a constant value at large $J/E_K$ consistent with the formulas:
\begin{subequations}
\begin{align}
    \rho^{\text{R,orb}}_\| &= 2\rho^{\text{R,orb}}_\perp =\chi_s \alpha_{\text{FS}} \frac{v_F G}{4\pi c} \label{eq:rholargeJorb}\\
    \rho^{\text{R,spin}}_\| &= 2\rho^{\text{R,spin}}_\perp = \chi_s \alpha_{\text{FS}}  \frac{g_sE_FG}{4\pi  m_e  v_F c} , \label{eq:rholargeJspin}
\end{align}
\label{eq:rholargeJ}
\end{subequations}
where $\alpha_{\text{FS}} = e^2 /2\varepsilon_0 h c$ is the fine structure constant~\cite{supp}. Rotatory power comparable to quartz is obtained even at the modest Fermi energy $E_F \sim 50$ meV. Here, two pairs of Weyl nodes related by $\mathcal{T}$-symmetry are assumed and the numerical values obtained for a single pair are doubled in the plots of Fig.~\ref{fig:numerics}. Internode scattering is expected to be negligible in the physically relevant regime of small $J$.
The behavior of the electronic spectrum at large $J$ deserves some comment. As shown schematically in the inset of Fig.~\ref{fig:numerics}b, one effect of the spin texture is to move the energy of one Weyl point upward and the other one downward, creating an effective energy splitting $\Delta \varepsilon\sim 2J$ due to the $\mathcal{P}$-breaking in the spin texture. A more accurate description of the energy spectrum at large $J$ is in terms of Landau level formation arising from the effective magnetic field generated by the coupling to the spiral spin texture, from which the above formulas can be derived~\cite{supp}. 
\par 
Finally, we plot the rotatory power for non-relativistic electrons and Weyl electrons coupled to the multi-$q$ texture in Fig.~\ref{fig:numerics}a,c, displaying excellent agreement with analytical results. 
At higher $J/E_K$ and $E_F$, a number of bands and Fermi pockets may contribute to the rotatory power, and discontinuities arise at intermediate $J/E_K$ due to changes in Fermi surface topology~\cite{supp}. While our examples are for one particular multi-$q$ spin texture, the existence of the GME extends to all $\mathcal{P}$-breaking metallic chiral magnets, including those hosting skyrmions, hedgehog-antihedgehog lattices, or any non-centrosymmetric spin motif.
\\

\textit{Discussion.} --- Through analytic and numerical model calculations we established that a finite, observable rotatory power is generic in metallic chiral magnets without an inversion center. Non-centrosymmetric metals such as MnSi and FeGe posessing the B20 crystal structure and spiral-spin ground state~\cite{Nagaosa2013Dec} satisfy the symmetry requirements to exhibit the GME. The same compounds host a skyrmion crystal phase upon application of mild magnetic fields, and the transition can be picked up through the sudden change in the rotation angle of light passing through. Kerr rotation of light reflecting off such skyrmion crystals has been observed recently in Gd$_2$PdSi$_3$~\cite{tokura23} and CrVI$_6$~\cite{Li24}. Instead of a reflection experiment, a transmission-type optical probe such as performed in \cite{Masuda2021Apr} should reveal the existence of the GME (but no Faraday effect) in the spiral phases of such compounds.
\par 
The helical Weyl semimetals SmAlSi~\cite{tafti23} and EuCuAs~\cite{Soh2024Jan} may be viable platforms to observe the proposed GME among the relativistic metals. The large optical Hall conductivity in the chiral magnet MnGe was attributed to the existence of Weyl nodes~\cite{Hayashi2021Oct} near the Fermi level arising from the multi-$q$ hedgehog spin structure. According to our theory, this material should exhibit the GME at low frequencies. EuIn$_2$As$_2$ has attracted considerable attention in recent years as an axion insulator candidate~\cite{dai19}. With the observation of a helical spin texture~\cite{riberolles21} and metallic conduction~\cite{yamashita22} rather than insulating behavior, however, this material might possess the ingredients to exhibit the GME. 
\par
Centrosymmetric skyrmion materials or helimagnets such as Gd$_2$PdSi$_3$, Gd$_3$Ru$_4$Al$_{12}$, and GdRu$_2$Si$_2$~\cite{kurumaji2019skyrmion,hirschberger2020high,hirschberger2019skyrmion,khanh2020nanometric,herpin1961etude,masuda2024room,moon1982neutron,PhysRevB.93.100405} could serve as ideal probes of the GME. In such systems, the GME can serve as a sharp indicator of non-centrosymmetric magnetic ordering due to its vanishing in the centrosymmetric parent state. Moreover, frustrated centrosymmetric magnets host spin textures of shorter typical periods than those arising from the Dzyaloshinskii–Moriya interaction ($\sim2$ nm vs. $\sim20$ nm) and Gd-based compounds can have a sizable Hunds coupling ($\sim 0.25$ eV~\cite{chen2025}), both resulting in a larger GME. We further discuss these candidate materials and provide estimates of the GME in \cite{supp}. 

The orbital contribution to the GME is substantial for relativistic electrons with their large intrinsic Berry curvature, while almost negligible for non-relativistic electrons. The spin contribution to the GME can be substantial for non-relativistic electrons at high density with large Fermi energy. Also noteworthy is that our calculation is not limited to Weyl electrons, but may well apply to Dirac metals with Weyl nodes of opposite chiralities centered about the same momentum. 
\par 
In practice, the GME can be isolated from the Faraday effect by measuring optical rotation along opposite directions $\pm \hat{\vec n}$ and taking the sum (GME) and difference (Faraday) of angles. Crucially, the GME is sensitive to spiral magnetic textures-- which are common in chiral magnets-- while the Faraday and Kerr effects are not~\cite{supp}. Moreover, the GME is symmetry-constrained and direction-dependent, making it a potential probe of the magnetic point-group symmetries and symmetry-breaking transitions. Overall, our findings open a new path to characterizing chiral magnets through optical means.



\acknowledgments
NP thanks Joel Moore for helpful discussions. JHH was supported by the National Research Foundation of Korea (NRF) grant funded by the Korea government (MSTI) (No. RS-2024-00410027).  TP is supported by the Air Force Office of Scientific Research (AFOSR) (award FA9550-22-1-0432).  LB is supported by the NSF CMMT program under Grants No. DMR-2419871, and the Simons Collaboration on Ultra-Quantum Matter, which is a grant from the Simons Foundation (Grant No. 651440). NP thanks the Kavli Institute for Theoretical Physics
for hospitality; KITP is supported in part by the Heising-Simons Foundation, the Simons Foundation, and the National Science Foundation under Grant PHY-2309135.

\par

\bibliography{ref}

\begin{thebibliography}{51}%
\makeatletter
\providecommand \@ifxundefined [1]{%
 \@ifx{#1\undefined}
}%
\providecommand \@ifnum [1]{%
 \ifnum #1\expandafter \@firstoftwo
 \else \expandafter \@secondoftwo
 \fi
}%
\providecommand \@ifx [1]{%
 \ifx #1\expandafter \@firstoftwo
 \else \expandafter \@secondoftwo
 \fi
}%
\providecommand \natexlab [1]{#1}%
\providecommand \enquote  [1]{``#1''}%
\providecommand \bibnamefont  [1]{#1}%
\providecommand \bibfnamefont [1]{#1}%
\providecommand \citenamefont [1]{#1}%
\providecommand \href@noop [0]{\@secondoftwo}%
\providecommand \href [0]{\begingroup \@sanitize@url \@href}%
\providecommand \@href[1]{\@@startlink{#1}\@@href}%
\providecommand \@@href[1]{\endgroup#1\@@endlink}%
\providecommand \@sanitize@url [0]{\catcode `\\12\catcode `\$12\catcode `\&12\catcode `\#12\catcode `\^12\catcode `\_12\catcode `\%12\relax}%
\providecommand \@@startlink[1]{}%
\providecommand \@@endlink[0]{}%
\providecommand \url  [0]{\begingroup\@sanitize@url \@url }%
\providecommand \@url [1]{\endgroup\@href {#1}{\urlprefix }}%
\providecommand \urlprefix  [0]{URL }%
\providecommand \Eprint [0]{\href }%
\providecommand \doibase [0]{https://doi.org/}%
\providecommand \selectlanguage [0]{\@gobble}%
\providecommand \bibinfo  [0]{\@secondoftwo}%
\providecommand \bibfield  [0]{\@secondoftwo}%
\providecommand \translation [1]{[#1]}%
\providecommand \BibitemOpen [0]{}%
\providecommand \bibitemStop [0]{}%
\providecommand \bibitemNoStop [0]{.\EOS\space}%
\providecommand \EOS [0]{\spacefactor3000\relax}%
\providecommand \BibitemShut  [1]{\csname bibitem#1\endcsname}%
\let\auto@bib@innerbib\@empty
\bibitem [{\citenamefont {Landau}\ and\ \citenamefont {Lifshitz}(1984)}]{Landau1984}%
  \BibitemOpen
  \bibfield  {author} {\bibinfo {author} {\bibfnamefont {L.~D.}\ \bibnamefont {Landau}}\ and\ \bibinfo {author} {\bibfnamefont {E.~M.}\ \bibnamefont {Lifshitz}},\ }\href@noop {} {\emph {\bibinfo {title} {{Electrodynamics of Continuous Media}}}}\ (\bibinfo  {publisher} {Pergamon},\ \bibinfo {address} {Oxford, England, UK},\ \bibinfo {year} {1984})\BibitemShut {NoStop}%
\bibitem [{\citenamefont {Agranovich}\ and\ \citenamefont {Ginzburg}()}]{Agranovich}%
  \BibitemOpen
  \bibfield  {author} {\bibinfo {author} {\bibfnamefont {V.~M.}\ \bibnamefont {Agranovich}}\ and\ \bibinfo {author} {\bibfnamefont {V.}~\bibnamefont {Ginzburg}},\ }\href {https://link.springer.com/book/10.1007/978-3-662-02406-5} {\emph {\bibinfo {title} {{Crystal Optics with Spatial Dispersion, and Excitons}}}}\ (\bibinfo  {publisher} {Springer},\ \bibinfo {address} {Berlin, Germany})\BibitemShut {NoStop}%
\bibitem [{\citenamefont {Jerphagnon}\ and\ \citenamefont {Chemla}(1976)}]{Jerphagnon1976Aug}%
  \BibitemOpen
  \bibfield  {author} {\bibinfo {author} {\bibfnamefont {J.}~\bibnamefont {Jerphagnon}}\ and\ \bibinfo {author} {\bibfnamefont {D.~S.}\ \bibnamefont {Chemla}},\ }\bibfield  {title} {\bibinfo {title} {{Optical activity of crystals}},\ }\href {https://doi.org/10.1063/1.433207} {\bibfield  {journal} {\bibinfo  {journal} {J. Chem. Phys.}\ }\textbf {\bibinfo {volume} {65}},\ \bibinfo {pages} {1522} (\bibinfo {year} {1976})}\BibitemShut {NoStop}%
\bibitem [{\citenamefont {Cheong}(2019)}]{Cheong2019Oct}%
  \BibitemOpen
  \bibfield  {author} {\bibinfo {author} {\bibfnamefont {S.-W.}\ \bibnamefont {Cheong}},\ }\bibfield  {title} {\bibinfo {title} {{SOS: symmetry-operational similarity}},\ }\href {https://doi.org/10.1038/s41535-019-0193-9} {\bibfield  {journal} {\bibinfo  {journal} {npj Quantum Mater.}\ }\textbf {\bibinfo {volume} {4}},\ \bibinfo {pages} {1} (\bibinfo {year} {2019})}\BibitemShut {NoStop}%
\bibitem [{\citenamefont {Arago}(1811)}]{arago1811}%
  \BibitemOpen
  \bibfield  {author} {\bibinfo {author} {\bibfnamefont {F.}~\bibnamefont {Arago}},\ }\href@noop {} {\emph {\bibinfo {title} {M{\'e}moire sur une modification remarquable qu' {\'e}prouvent les rayons lumineux dans leur passage {\`a} travers certains corps diaphanes, \& sur quelques autres ph{\'e}nom{\`e}nes d'optique}}},\ M{\'e}moire de la Classe des sciences math{\'e}matiques \& physiques de l'Acad{\'e}mie des Sciences\ (\bibinfo {year} {1811})\BibitemShut {NoStop}%
\bibitem [{\citenamefont {Fried}(2014)}]{fried14}%
  \BibitemOpen
  \bibfield  {author} {\bibinfo {author} {\bibfnamefont {A.~D.}\ \bibnamefont {Fried}},\ }\bibfield  {title} {\bibinfo {title} {Relationship of time-reversal symmetry breaking to optical kerr rotation},\ }\href {https://doi.org/10.1103/PhysRevB.90.121112} {\bibfield  {journal} {\bibinfo  {journal} {Phys. Rev. B}\ }\textbf {\bibinfo {volume} {90}},\ \bibinfo {pages} {121112} (\bibinfo {year} {2014})}\BibitemShut {NoStop}%
\bibitem [{\citenamefont {Hayashi}\ \emph {et~al.}(2021{\natexlab{a}})\citenamefont {Hayashi}, \citenamefont {Okamura}, \citenamefont {Kanazawa}, \citenamefont {Yu}, \citenamefont {Koretsune}, \citenamefont {Arita}, \citenamefont {Tsukazaki}, \citenamefont {Ichikawa}, \citenamefont {Kawasaki}, \citenamefont {Tokura},\ and\ \citenamefont {Takahashi}}]{MnGe-NatComm21}%
  \BibitemOpen
  \bibfield  {author} {\bibinfo {author} {\bibfnamefont {Y.}~\bibnamefont {Hayashi}}, \bibinfo {author} {\bibfnamefont {Y.}~\bibnamefont {Okamura}}, \bibinfo {author} {\bibfnamefont {N.}~\bibnamefont {Kanazawa}}, \bibinfo {author} {\bibfnamefont {T.}~\bibnamefont {Yu}}, \bibinfo {author} {\bibfnamefont {T.}~\bibnamefont {Koretsune}}, \bibinfo {author} {\bibfnamefont {R.}~\bibnamefont {Arita}}, \bibinfo {author} {\bibfnamefont {A.}~\bibnamefont {Tsukazaki}}, \bibinfo {author} {\bibfnamefont {M.}~\bibnamefont {Ichikawa}}, \bibinfo {author} {\bibfnamefont {M.}~\bibnamefont {Kawasaki}}, \bibinfo {author} {\bibfnamefont {Y.}~\bibnamefont {Tokura}},\ and\ \bibinfo {author} {\bibfnamefont {Y.}~\bibnamefont {Takahashi}},\ }\bibfield  {title} {\bibinfo {title} {Magneto-optical spectroscopy on weyl nodes for anomalous and topological hall effects in chiral mnge},\ }\href {https://doi.org/10.1038/s41467-021-25276-1} {\bibfield  {journal} {\bibinfo  {journal} {Nature Communications}\ }\textbf {\bibinfo {volume} {12}},\
  \bibinfo {pages} {5974} (\bibinfo {year} {2021}{\natexlab{a}})}\BibitemShut {NoStop}%
\bibitem [{\citenamefont {Barron}(2009)}]{barron09}%
  \BibitemOpen
  \bibfield  {author} {\bibinfo {author} {\bibfnamefont {L.~D.}\ \bibnamefont {Barron}},\ }\href@noop {} {\emph {\bibinfo {title} {Molecular Light Scattering and Optical Activity}}},\ \bibinfo {edition} {2nd}\ ed.\ (\bibinfo  {publisher} {Cambridge University Press},\ \bibinfo {address} {Cambridge, UK},\ \bibinfo {year} {2009})\BibitemShut {NoStop}%
\bibitem [{\citenamefont {Ma}\ and\ \citenamefont {Pesin}(2015)}]{pesin15}%
  \BibitemOpen
  \bibfield  {author} {\bibinfo {author} {\bibfnamefont {J.}~\bibnamefont {Ma}}\ and\ \bibinfo {author} {\bibfnamefont {D.~A.}\ \bibnamefont {Pesin}},\ }\bibfield  {title} {\bibinfo {title} {Chiral magnetic effect and natural optical activity in metals with or without weyl points},\ }\href {https://doi.org/10.1103/PhysRevB.92.235205} {\bibfield  {journal} {\bibinfo  {journal} {Phys. Rev. B}\ }\textbf {\bibinfo {volume} {92}},\ \bibinfo {pages} {235205} (\bibinfo {year} {2015})}\BibitemShut {NoStop}%
\bibitem [{\citenamefont {Goswami}\ \emph {et~al.}(2015)\citenamefont {Goswami}, \citenamefont {Sharma},\ and\ \citenamefont {Tewari}}]{Goswami2015Oct}%
  \BibitemOpen
  \bibfield  {author} {\bibinfo {author} {\bibfnamefont {P.}~\bibnamefont {Goswami}}, \bibinfo {author} {\bibfnamefont {G.}~\bibnamefont {Sharma}},\ and\ \bibinfo {author} {\bibfnamefont {S.}~\bibnamefont {Tewari}},\ }\bibfield  {title} {\bibinfo {title} {{Optical activity as a test for dynamic chiral magnetic effect of Weyl semimetals}},\ }\href {https://doi.org/10.1103/PhysRevB.92.161110} {\bibfield  {journal} {\bibinfo  {journal} {Phys. Rev. B}\ }\textbf {\bibinfo {volume} {92}},\ \bibinfo {pages} {161110} (\bibinfo {year} {2015})}\BibitemShut {NoStop}%
\bibitem [{\citenamefont {Zhong}\ \emph {et~al.}(2016)\citenamefont {Zhong}, \citenamefont {Moore},\ and\ \citenamefont {Souza}}]{souza16}%
  \BibitemOpen
  \bibfield  {author} {\bibinfo {author} {\bibfnamefont {S.}~\bibnamefont {Zhong}}, \bibinfo {author} {\bibfnamefont {J.~E.}\ \bibnamefont {Moore}},\ and\ \bibinfo {author} {\bibfnamefont {I.}~\bibnamefont {Souza}},\ }\bibfield  {title} {\bibinfo {title} {Gyrotropic magnetic effect and the magnetic moment on the fermi surface},\ }\href {https://doi.org/10.1103/PhysRevLett.116.077201} {\bibfield  {journal} {\bibinfo  {journal} {Phys. Rev. Lett.}\ }\textbf {\bibinfo {volume} {116}},\ \bibinfo {pages} {077201} (\bibinfo {year} {2016})}\BibitemShut {NoStop}%
\bibitem [{\citenamefont {Shalygin}\ \emph {et~al.}(2012)\citenamefont {Shalygin}, \citenamefont {Sofronov}, \citenamefont {Vorob’ev},\ and\ \citenamefont {Farbshtein}}]{shalygin2012current}%
  \BibitemOpen
  \bibfield  {author} {\bibinfo {author} {\bibfnamefont {V.}~\bibnamefont {Shalygin}}, \bibinfo {author} {\bibfnamefont {A.}~\bibnamefont {Sofronov}}, \bibinfo {author} {\bibfnamefont {L.}~\bibnamefont {Vorob’ev}},\ and\ \bibinfo {author} {\bibfnamefont {I.}~\bibnamefont {Farbshtein}},\ }\bibfield  {title} {\bibinfo {title} {Current-induced spin polarization of holes in tellurium},\ }\href@noop {} {\bibfield  {journal} {\bibinfo  {journal} {Physics of the Solid State}\ }\textbf {\bibinfo {volume} {54}},\ \bibinfo {pages} {2362} (\bibinfo {year} {2012})}\BibitemShut {NoStop}%
\bibitem [{\citenamefont {Tsirkin}\ \emph {et~al.}(2018)\citenamefont {Tsirkin}, \citenamefont {Puente},\ and\ \citenamefont {Souza}}]{Tsirkin2018Jan}%
  \BibitemOpen
  \bibfield  {author} {\bibinfo {author} {\bibfnamefont {S.~S.}\ \bibnamefont {Tsirkin}}, \bibinfo {author} {\bibfnamefont {P.~A.}\ \bibnamefont {Puente}},\ and\ \bibinfo {author} {\bibfnamefont {I.}~\bibnamefont {Souza}},\ }\bibfield  {title} {\bibinfo {title} {{Gyrotropic effects in trigonal tellurium studied from first principles}},\ }\href {https://doi.org/10.1103/PhysRevB.97.035158} {\bibfield  {journal} {\bibinfo  {journal} {Phys. Rev. B}\ }\textbf {\bibinfo {volume} {97}},\ \bibinfo {pages} {035158} (\bibinfo {year} {2018})}\BibitemShut {NoStop}%
\bibitem [{\citenamefont {Cheong}\ and\ \citenamefont {Xu}(2022)}]{Cheong2022Apr}%
  \BibitemOpen
  \bibfield  {author} {\bibinfo {author} {\bibfnamefont {S.-W.}\ \bibnamefont {Cheong}}\ and\ \bibinfo {author} {\bibfnamefont {X.}~\bibnamefont {Xu}},\ }\bibfield  {title} {\bibinfo {title} {{Magnetic chirality}},\ }\href {https://doi.org/10.1038/s41535-022-00447-5} {\bibfield  {journal} {\bibinfo  {journal} {npj Quantum Mater.}\ }\textbf {\bibinfo {volume} {7}},\ \bibinfo {pages} {1} (\bibinfo {year} {2022})}\BibitemShut {NoStop}%
\bibitem [{\citenamefont {Nagaosa}\ and\ \citenamefont {Tokura}(2013)}]{Nagaosa2013Dec}%
  \BibitemOpen
  \bibfield  {author} {\bibinfo {author} {\bibfnamefont {N.}~\bibnamefont {Nagaosa}}\ and\ \bibinfo {author} {\bibfnamefont {Y.}~\bibnamefont {Tokura}},\ }\bibfield  {title} {\bibinfo {title} {{Topological properties and dynamics of magnetic skyrmions}},\ }\href {https://doi.org/10.1038/nnano.2013.243} {\bibfield  {journal} {\bibinfo  {journal} {Nat. Nanotechnol.}\ }\textbf {\bibinfo {volume} {8}},\ \bibinfo {pages} {899} (\bibinfo {year} {2013})}\BibitemShut {NoStop}%
\bibitem [{\citenamefont {Fert}\ \emph {et~al.}(2017)\citenamefont {Fert}, \citenamefont {Reyren},\ and\ \citenamefont {Cros}}]{Fert2017}%
  \BibitemOpen
  \bibfield  {author} {\bibinfo {author} {\bibfnamefont {A.}~\bibnamefont {Fert}}, \bibinfo {author} {\bibfnamefont {N.}~\bibnamefont {Reyren}},\ and\ \bibinfo {author} {\bibfnamefont {V.}~\bibnamefont {Cros}},\ }\bibfield  {title} {\bibinfo {title} {Magnetic skyrmions: advances in physics and potential applications},\ }\href {https://doi.org/10.1038/natrevmats.2017.31} {\bibfield  {journal} {\bibinfo  {journal} {Nature Reviews Materials}\ }\textbf {\bibinfo {volume} {2}},\ \bibinfo {pages} {17031} (\bibinfo {year} {2017})}\BibitemShut {NoStop}%
\bibitem [{\citenamefont {Han}(2017)}]{Han2017}%
  \BibitemOpen
  \bibfield  {author} {\bibinfo {author} {\bibfnamefont {J.~H.}\ \bibnamefont {Han}},\ }\href {https://link.springer.com/book/10.1007/978-3-319-69246-3} {\emph {\bibinfo {title} {{Skyrmions in Condensed Matter}}}}\ (\bibinfo  {publisher} {Springer International Publishing},\ \bibinfo {address} {Cham, Switzerland},\ \bibinfo {year} {2017})\BibitemShut {NoStop}%
\bibitem [{\citenamefont {Yang}\ \emph {et~al.}(2021)\citenamefont {Yang}, \citenamefont {Naaman}, \citenamefont {Paltiel},\ and\ \citenamefont {Parkin}}]{Yang2021May}%
  \BibitemOpen
  \bibfield  {author} {\bibinfo {author} {\bibfnamefont {S.-H.}\ \bibnamefont {Yang}}, \bibinfo {author} {\bibfnamefont {R.}~\bibnamefont {Naaman}}, \bibinfo {author} {\bibfnamefont {Y.}~\bibnamefont {Paltiel}},\ and\ \bibinfo {author} {\bibfnamefont {S.~S.~P.}\ \bibnamefont {Parkin}},\ }\bibfield  {title} {\bibinfo {title} {{Chiral spintronics}},\ }\href {https://doi.org/10.1038/s42254-021-00302-9} {\bibfield  {journal} {\bibinfo  {journal} {Nat. Rev. Phys.}\ }\textbf {\bibinfo {volume} {3}},\ \bibinfo {pages} {328} (\bibinfo {year} {2021})}\BibitemShut {NoStop}%
\bibitem [{\citenamefont {Tokura}\ and\ \citenamefont {Kanazawa}(2020)}]{Tokura2020Nov}%
  \BibitemOpen
  \bibfield  {author} {\bibinfo {author} {\bibfnamefont {Y.}~\bibnamefont {Tokura}}\ and\ \bibinfo {author} {\bibfnamefont {N.}~\bibnamefont {Kanazawa}},\ }\bibfield  {title} {\bibinfo {title} {{Magnetic Skyrmion Materials}},\ }\bibfield  {journal} {\bibinfo  {journal} {Chem. Rev.}\ }\href {https://doi.org/10.1021/acs.chemrev.0c00297} {10.1021/acs.chemrev.0c00297} (\bibinfo {year} {2020})\BibitemShut {NoStop}%
\bibitem [{\citenamefont {Masuda}\ \emph {et~al.}(2021)\citenamefont {Masuda}, \citenamefont {Kaneko}, \citenamefont {Tokura},\ and\ \citenamefont {Takahashi}}]{Masuda2021Apr}%
  \BibitemOpen
  \bibfield  {author} {\bibinfo {author} {\bibfnamefont {R.}~\bibnamefont {Masuda}}, \bibinfo {author} {\bibfnamefont {Y.}~\bibnamefont {Kaneko}}, \bibinfo {author} {\bibfnamefont {Y.}~\bibnamefont {Tokura}},\ and\ \bibinfo {author} {\bibfnamefont {Y.}~\bibnamefont {Takahashi}},\ }\bibfield  {title} {\bibinfo {title} {{Electric field control of natural optical activity in a multiferroic helimagnet}},\ }\href {https://doi.org/10.1126/science.aaz4312} {\bibfield  {journal} {\bibinfo  {journal} {Science}\ }\textbf {\bibinfo {volume} {372}},\ \bibinfo {pages} {496} (\bibinfo {year} {2021})}\BibitemShut {NoStop}%
\bibitem [{\citenamefont {Flicker}\ \emph {et~al.}(2018)\citenamefont {Flicker}, \citenamefont {de~Juan}, \citenamefont {Bradlyn}, \citenamefont {Morimoto}, \citenamefont {Vergniory},\ and\ \citenamefont {Grushin}}]{grushin18}%
  \BibitemOpen
  \bibfield  {author} {\bibinfo {author} {\bibfnamefont {F.}~\bibnamefont {Flicker}}, \bibinfo {author} {\bibfnamefont {F.}~\bibnamefont {de~Juan}}, \bibinfo {author} {\bibfnamefont {B.}~\bibnamefont {Bradlyn}}, \bibinfo {author} {\bibfnamefont {T.}~\bibnamefont {Morimoto}}, \bibinfo {author} {\bibfnamefont {M.~G.}\ \bibnamefont {Vergniory}},\ and\ \bibinfo {author} {\bibfnamefont {A.~G.}\ \bibnamefont {Grushin}},\ }\bibfield  {title} {\bibinfo {title} {Chiral optical response of multifold fermions},\ }\href {https://doi.org/10.1103/PhysRevB.98.155145} {\bibfield  {journal} {\bibinfo  {journal} {Phys. Rev. B}\ }\textbf {\bibinfo {volume} {98}},\ \bibinfo {pages} {155145} (\bibinfo {year} {2018})}\BibitemShut {NoStop}%
\bibitem [{\citenamefont {Wang}\ \emph {et~al.}(2019)\citenamefont {Wang}, \citenamefont {Lian},\ and\ \citenamefont {Zhang}}]{Wang2019Dec}%
  \BibitemOpen
  \bibfield  {author} {\bibinfo {author} {\bibfnamefont {J.}~\bibnamefont {Wang}}, \bibinfo {author} {\bibfnamefont {B.}~\bibnamefont {Lian}},\ and\ \bibinfo {author} {\bibfnamefont {S.-C.}\ \bibnamefont {Zhang}},\ }\bibfield  {title} {\bibinfo {title} {{Generation of Spin Currents by Magnetic Field in {$\mathscr{T}$}- and {$\mathscr{P}$}-Broken Materials}},\ }\href {https://doi.org/10.1142/S2010324719400137} {\bibfield  {journal} {\bibinfo  {journal} {SPIN}\ }\textbf {\bibinfo {volume} {09}},\ \bibinfo {pages} {1940013} (\bibinfo {year} {2019})}\BibitemShut {NoStop}%
\bibitem [{\citenamefont {Wang}\ \emph {et~al.}(2020{\natexlab{a}})\citenamefont {Wang}, \citenamefont {Morimoto},\ and\ \citenamefont {Moore}}]{Wang2020May}%
  \BibitemOpen
  \bibfield  {author} {\bibinfo {author} {\bibfnamefont {Y.-Q.}\ \bibnamefont {Wang}}, \bibinfo {author} {\bibfnamefont {T.}~\bibnamefont {Morimoto}},\ and\ \bibinfo {author} {\bibfnamefont {J.~E.}\ \bibnamefont {Moore}},\ }\bibfield  {title} {\bibinfo {title} {{Optical rotation in thin chiral/twisted materials and the gyrotropic magnetic effect}},\ }\href {https://doi.org/10.1103/PhysRevB.101.174419} {\bibfield  {journal} {\bibinfo  {journal} {Phys. Rev. B}\ }\textbf {\bibinfo {volume} {101}},\ \bibinfo {pages} {174419} (\bibinfo {year} {2020}{\natexlab{a}})}\BibitemShut {NoStop}%
\bibitem [{\citenamefont {Son}\ and\ \citenamefont {Yamamoto}(2012)}]{son-yamamoto}%
  \BibitemOpen
  \bibfield  {author} {\bibinfo {author} {\bibfnamefont {D.~T.}\ \bibnamefont {Son}}\ and\ \bibinfo {author} {\bibfnamefont {N.}~\bibnamefont {Yamamoto}},\ }\bibfield  {title} {\bibinfo {title} {Berry curvature, triangle anomalies, and the chiral magnetic effect in fermi liquids},\ }\href {https://doi.org/10.1103/PhysRevLett.109.181602} {\bibfield  {journal} {\bibinfo  {journal} {Phys. Rev. Lett.}\ }\textbf {\bibinfo {volume} {109}},\ \bibinfo {pages} {181602} (\bibinfo {year} {2012})}\BibitemShut {NoStop}%
\bibitem [{\citenamefont {Zyuzin}\ and\ \citenamefont {Burkov}(2012)}]{burkov12}%
  \BibitemOpen
  \bibfield  {author} {\bibinfo {author} {\bibfnamefont {A.~A.}\ \bibnamefont {Zyuzin}}\ and\ \bibinfo {author} {\bibfnamefont {A.~A.}\ \bibnamefont {Burkov}},\ }\bibfield  {title} {\bibinfo {title} {Topological response in weyl semimetals and the chiral anomaly},\ }\href {https://doi.org/10.1103/PhysRevB.86.115133} {\bibfield  {journal} {\bibinfo  {journal} {Phys. Rev. B}\ }\textbf {\bibinfo {volume} {86}},\ \bibinfo {pages} {115133} (\bibinfo {year} {2012})}\BibitemShut {NoStop}%
\bibitem [{\citenamefont {Son}\ and\ \citenamefont {Spivak}(2013)}]{son-spivak}%
  \BibitemOpen
  \bibfield  {author} {\bibinfo {author} {\bibfnamefont {D.~T.}\ \bibnamefont {Son}}\ and\ \bibinfo {author} {\bibfnamefont {B.~Z.}\ \bibnamefont {Spivak}},\ }\bibfield  {title} {\bibinfo {title} {Chiral anomaly and classical negative magnetoresistance of weyl metals},\ }\href {https://doi.org/10.1103/PhysRevB.88.104412} {\bibfield  {journal} {\bibinfo  {journal} {Phys. Rev. B}\ }\textbf {\bibinfo {volume} {88}},\ \bibinfo {pages} {104412} (\bibinfo {year} {2013})}\BibitemShut {NoStop}%
\bibitem [{\citenamefont {Jian-Hui}\ \emph {et~al.}(2013)\citenamefont {Jian-Hui}, \citenamefont {Hua}, \citenamefont {Qian},\ and\ \citenamefont {Jun-Ren}}]{niu13}%
  \BibitemOpen
  \bibfield  {author} {\bibinfo {author} {\bibfnamefont {Z.}~\bibnamefont {Jian-Hui}}, \bibinfo {author} {\bibfnamefont {J.}~\bibnamefont {Hua}}, \bibinfo {author} {\bibfnamefont {N.}~\bibnamefont {Qian}},\ and\ \bibinfo {author} {\bibfnamefont {S.}~\bibnamefont {Jun-Ren}},\ }\bibfield  {title} {\bibinfo {title} {{Topological Invariants of Metals and the Related Physical Effects}},\ }\href {https://doi.org/10.1088/0256-307X/30/2/027101} {\bibfield  {journal} {\bibinfo  {journal} {Chin. Phys. Lett.}\ }\textbf {\bibinfo {volume} {30}},\ \bibinfo {pages} {027101} (\bibinfo {year} {2013})}\BibitemShut {NoStop}%
\bibitem [{\citenamefont {Newnham}(2005)}]{Newnham2005}%
  \BibitemOpen
  \bibfield  {author} {\bibinfo {author} {\bibfnamefont {R.~E.}\ \bibnamefont {Newnham}},\ }\href@noop {} {\emph {\bibinfo {title} {{Properties of Materials: Anisotropy, Symmetry, Structure}}}}\ (\bibinfo  {publisher} {OUP},\ \bibinfo {address} {Oxford, England, UK},\ \bibinfo {year} {2005})\BibitemShut {NoStop}%
\bibitem [{sup()}]{supp}%
  \BibitemOpen
  \href@noop {} {}\bibinfo {note} {See the Supplemental Material at [URL] for background theory on optical activity, details of perturbative and numerical calculations, the form of the mulit$-q$ ansatz, details on the single-particle spectrum, and material estimates, which includes Refs.~\cite{moore20,Park2011May,jackiw-rebbi,chen2025,kurumaji2019skyrmion,hirschberger2020high,paddison,hirschberger2019skyrmion,khanh2020nanometric,herpin1961etude,masuda2024room,moon1982neutron,PhysRevB.93.100405}.}\BibitemShut {Stop}%
\bibitem [{\citenamefont {Tanigaki}\ \emph {et~al.}(2015)\citenamefont {Tanigaki}, \citenamefont {Shibata}, \citenamefont {Kanazawa}, \citenamefont {Yu}, \citenamefont {Onose}, \citenamefont {Park}, \citenamefont {Shindo},\ and\ \citenamefont {Tokura}}]{MnGe-NanoLett15}%
  \BibitemOpen
  \bibfield  {author} {\bibinfo {author} {\bibfnamefont {T.}~\bibnamefont {Tanigaki}}, \bibinfo {author} {\bibfnamefont {K.}~\bibnamefont {Shibata}}, \bibinfo {author} {\bibfnamefont {N.}~\bibnamefont {Kanazawa}}, \bibinfo {author} {\bibfnamefont {X.}~\bibnamefont {Yu}}, \bibinfo {author} {\bibfnamefont {Y.}~\bibnamefont {Onose}}, \bibinfo {author} {\bibfnamefont {H.~S.}\ \bibnamefont {Park}}, \bibinfo {author} {\bibfnamefont {D.}~\bibnamefont {Shindo}},\ and\ \bibinfo {author} {\bibfnamefont {Y.}~\bibnamefont {Tokura}},\ }\bibfield  {title} {\bibinfo {title} {Real-space observation of short-period cubic lattice of skyrmions in mnge},\ }\href {https://doi.org/10.1021/acs.nanolett.5b02653} {\bibfield  {journal} {\bibinfo  {journal} {Nano Letters}\ }\textbf {\bibinfo {volume} {15}},\ \bibinfo {pages} {5438} (\bibinfo {year} {2015})}\BibitemShut {NoStop}%
\bibitem [{\citenamefont {Kato}\ \emph {et~al.}(2023)\citenamefont {Kato}, \citenamefont {Okamura}, \citenamefont {Hirschberger}, \citenamefont {Tokura},\ and\ \citenamefont {Takahashi}}]{tokura23}%
  \BibitemOpen
  \bibfield  {author} {\bibinfo {author} {\bibfnamefont {Y.~D.}\ \bibnamefont {Kato}}, \bibinfo {author} {\bibfnamefont {Y.}~\bibnamefont {Okamura}}, \bibinfo {author} {\bibfnamefont {M.}~\bibnamefont {Hirschberger}}, \bibinfo {author} {\bibfnamefont {Y.}~\bibnamefont {Tokura}},\ and\ \bibinfo {author} {\bibfnamefont {Y.}~\bibnamefont {Takahashi}},\ }\bibfield  {title} {\bibinfo {title} {Topological magneto-optical effect from skyrmion lattice},\ }\href {https://doi.org/10.1038/s41467-023-41203-y} {\bibfield  {journal} {\bibinfo  {journal} {Nature Communications}\ }\textbf {\bibinfo {volume} {14}},\ \bibinfo {pages} {5416} (\bibinfo {year} {2023})}\BibitemShut {NoStop}%
\bibitem [{\citenamefont {Li}\ \emph {et~al.}(2024)\citenamefont {Li}, \citenamefont {Liu}, \citenamefont {Zhang}, \citenamefont {Zhang}, \citenamefont {Zhang}, \citenamefont {Zhang}, \citenamefont {Meng}, \citenamefont {Hou}, \citenamefont {Li}, \citenamefont {Kang}, \citenamefont {Huang}, \citenamefont {Cao}, \citenamefont {Hou}, \citenamefont {Cui}, \citenamefont {Zhang}, \citenamefont {Min}, \citenamefont {Lu}, \citenamefont {Xu}, \citenamefont {Sheng}, \citenamefont {Xiang},\ and\ \citenamefont {Zhang}}]{Li24}%
  \BibitemOpen
  \bibfield  {author} {\bibinfo {author} {\bibfnamefont {X.}~\bibnamefont {Li}}, \bibinfo {author} {\bibfnamefont {C.}~\bibnamefont {Liu}}, \bibinfo {author} {\bibfnamefont {Y.}~\bibnamefont {Zhang}}, \bibinfo {author} {\bibfnamefont {S.}~\bibnamefont {Zhang}}, \bibinfo {author} {\bibfnamefont {H.}~\bibnamefont {Zhang}}, \bibinfo {author} {\bibfnamefont {Y.}~\bibnamefont {Zhang}}, \bibinfo {author} {\bibfnamefont {W.}~\bibnamefont {Meng}}, \bibinfo {author} {\bibfnamefont {D.}~\bibnamefont {Hou}}, \bibinfo {author} {\bibfnamefont {T.}~\bibnamefont {Li}}, \bibinfo {author} {\bibfnamefont {C.}~\bibnamefont {Kang}}, \bibinfo {author} {\bibfnamefont {F.}~\bibnamefont {Huang}}, \bibinfo {author} {\bibfnamefont {R.}~\bibnamefont {Cao}}, \bibinfo {author} {\bibfnamefont {D.}~\bibnamefont {Hou}}, \bibinfo {author} {\bibfnamefont {P.}~\bibnamefont {Cui}}, \bibinfo {author} {\bibfnamefont {W.}~\bibnamefont {Zhang}}, \bibinfo {author} {\bibfnamefont {T.}~\bibnamefont {Min}}, \bibinfo {author} {\bibfnamefont
  {Q.}~\bibnamefont {Lu}}, \bibinfo {author} {\bibfnamefont {X.}~\bibnamefont {Xu}}, \bibinfo {author} {\bibfnamefont {Z.}~\bibnamefont {Sheng}}, \bibinfo {author} {\bibfnamefont {B.}~\bibnamefont {Xiang}},\ and\ \bibinfo {author} {\bibfnamefont {Z.}~\bibnamefont {Zhang}},\ }\bibfield  {title} {\bibinfo {title} {Topological kerr effects in two-dimensional magnets with broken inversion symmetry},\ }\href {https://doi.org/10.1038/s41567-024-02465-5} {\bibfield  {journal} {\bibinfo  {journal} {Nature Physics}\ }\textbf {\bibinfo {volume} {20}},\ \bibinfo {pages} {1145} (\bibinfo {year} {2024})}\BibitemShut {NoStop}%
\bibitem [{\citenamefont {Yao}\ \emph {et~al.}(2023)\citenamefont {Yao}, \citenamefont {Gaudet}, \citenamefont {Verma}, \citenamefont {Graf}, \citenamefont {Yang}, \citenamefont {Bahrami}, \citenamefont {Zhang}, \citenamefont {Aczel}, \citenamefont {Subedi}, \citenamefont {Torchinsky}, \citenamefont {Sun}, \citenamefont {Bansil}, \citenamefont {Huang}, \citenamefont {Singh}, \citenamefont {Blaha}, \citenamefont {Nikoli\ifmmode~\acute{c}\else \'{c}\fi{}},\ and\ \citenamefont {Tafti}}]{tafti23}%
  \BibitemOpen
  \bibfield  {author} {\bibinfo {author} {\bibfnamefont {X.}~\bibnamefont {Yao}}, \bibinfo {author} {\bibfnamefont {J.}~\bibnamefont {Gaudet}}, \bibinfo {author} {\bibfnamefont {R.}~\bibnamefont {Verma}}, \bibinfo {author} {\bibfnamefont {D.~E.}\ \bibnamefont {Graf}}, \bibinfo {author} {\bibfnamefont {H.-Y.}\ \bibnamefont {Yang}}, \bibinfo {author} {\bibfnamefont {F.}~\bibnamefont {Bahrami}}, \bibinfo {author} {\bibfnamefont {R.}~\bibnamefont {Zhang}}, \bibinfo {author} {\bibfnamefont {A.~A.}\ \bibnamefont {Aczel}}, \bibinfo {author} {\bibfnamefont {S.}~\bibnamefont {Subedi}}, \bibinfo {author} {\bibfnamefont {D.~H.}\ \bibnamefont {Torchinsky}}, \bibinfo {author} {\bibfnamefont {J.}~\bibnamefont {Sun}}, \bibinfo {author} {\bibfnamefont {A.}~\bibnamefont {Bansil}}, \bibinfo {author} {\bibfnamefont {S.-M.}\ \bibnamefont {Huang}}, \bibinfo {author} {\bibfnamefont {B.}~\bibnamefont {Singh}}, \bibinfo {author} {\bibfnamefont {P.}~\bibnamefont {Blaha}}, \bibinfo {author} {\bibfnamefont {P.}~\bibnamefont
  {Nikoli\ifmmode~\acute{c}\else \'{c}\fi{}}},\ and\ \bibinfo {author} {\bibfnamefont {F.}~\bibnamefont {Tafti}},\ }\bibfield  {title} {\bibinfo {title} {Large topological hall effect and spiral magnetic order in the weyl semimetal smalsi},\ }\href {https://doi.org/10.1103/PhysRevX.13.011035} {\bibfield  {journal} {\bibinfo  {journal} {Phys. Rev. X}\ }\textbf {\bibinfo {volume} {13}},\ \bibinfo {pages} {011035} (\bibinfo {year} {2023})}\BibitemShut {NoStop}%
\bibitem [{\citenamefont {Soh}\ \emph {et~al.}(2024)\citenamefont {Soh}, \citenamefont {S{\ifmmode\acute{a}\else\'{a}\fi}nchez-Ram{\ifmmode\acute{\imath}\else\'{\i}\fi}rez}, \citenamefont {Yang}, \citenamefont {Sun}, \citenamefont {Zivkovic}, \citenamefont {Rodr{\ifmmode\acute{\imath}\else\'{\i}\fi}guez-Velamaz{\ifmmode\acute{a}\else\'{a}\fi}n}, \citenamefont {Fabelo}, \citenamefont {Stunault}, \citenamefont {Bombardi}, \citenamefont {Balz}, \citenamefont {Le}, \citenamefont {Walker}, \citenamefont {Dil}, \citenamefont {Prabhakaran}, \citenamefont {R{\o}nnow}, \citenamefont {de~Juan}, \citenamefont {Vergniory},\ and\ \citenamefont {Boothroyd}}]{Soh2024Jan}%
  \BibitemOpen
  \bibfield  {author} {\bibinfo {author} {\bibfnamefont {J.-R.}\ \bibnamefont {Soh}}, \bibinfo {author} {\bibfnamefont {I.}~\bibnamefont {S{\ifmmode\acute{a}\else\'{a}\fi}nchez-Ram{\ifmmode\acute{\imath}\else\'{\i}\fi}rez}}, \bibinfo {author} {\bibfnamefont {X.}~\bibnamefont {Yang}}, \bibinfo {author} {\bibfnamefont {J.}~\bibnamefont {Sun}}, \bibinfo {author} {\bibfnamefont {I.}~\bibnamefont {Zivkovic}}, \bibinfo {author} {\bibfnamefont {J.~A.}\ \bibnamefont {Rodr{\ifmmode\acute{\imath}\else\'{\i}\fi}guez-Velamaz{\ifmmode\acute{a}\else\'{a}\fi}n}}, \bibinfo {author} {\bibfnamefont {O.}~\bibnamefont {Fabelo}}, \bibinfo {author} {\bibfnamefont {A.}~\bibnamefont {Stunault}}, \bibinfo {author} {\bibfnamefont {A.}~\bibnamefont {Bombardi}}, \bibinfo {author} {\bibfnamefont {C.}~\bibnamefont {Balz}}, \bibinfo {author} {\bibfnamefont {M.~D.}\ \bibnamefont {Le}}, \bibinfo {author} {\bibfnamefont {H.~C.}\ \bibnamefont {Walker}}, \bibinfo {author} {\bibfnamefont {J.~H.}\ \bibnamefont {Dil}}, \bibinfo {author}
  {\bibfnamefont {D.}~\bibnamefont {Prabhakaran}}, \bibinfo {author} {\bibfnamefont {H.~M.}\ \bibnamefont {R{\o}nnow}}, \bibinfo {author} {\bibfnamefont {F.}~\bibnamefont {de~Juan}}, \bibinfo {author} {\bibfnamefont {M.~G.}\ \bibnamefont {Vergniory}},\ and\ \bibinfo {author} {\bibfnamefont {A.~T.}\ \bibnamefont {Boothroyd}},\ }\bibfield  {title} {\bibinfo {title} {{Weyl metallic state induced by helical magnetic order}},\ }\href {https://doi.org/10.1038/s41535-023-00604-4} {\bibfield  {journal} {\bibinfo  {journal} {npj Quantum Mater.}\ }\textbf {\bibinfo {volume} {9}},\ \bibinfo {pages} {1} (\bibinfo {year} {2024})}\BibitemShut {NoStop}%
\bibitem [{\citenamefont {Hayashi}\ \emph {et~al.}(2021{\natexlab{b}})\citenamefont {Hayashi}, \citenamefont {Okamura}, \citenamefont {Kanazawa}, \citenamefont {Yu}, \citenamefont {Koretsune}, \citenamefont {Arita}, \citenamefont {Tsukazaki}, \citenamefont {Ichikawa}, \citenamefont {Kawasaki}, \citenamefont {Tokura},\ and\ \citenamefont {Takahashi}}]{Hayashi2021Oct}%
  \BibitemOpen
  \bibfield  {author} {\bibinfo {author} {\bibfnamefont {Y.}~\bibnamefont {Hayashi}}, \bibinfo {author} {\bibfnamefont {Y.}~\bibnamefont {Okamura}}, \bibinfo {author} {\bibfnamefont {N.}~\bibnamefont {Kanazawa}}, \bibinfo {author} {\bibfnamefont {T.}~\bibnamefont {Yu}}, \bibinfo {author} {\bibfnamefont {T.}~\bibnamefont {Koretsune}}, \bibinfo {author} {\bibfnamefont {R.}~\bibnamefont {Arita}}, \bibinfo {author} {\bibfnamefont {A.}~\bibnamefont {Tsukazaki}}, \bibinfo {author} {\bibfnamefont {M.}~\bibnamefont {Ichikawa}}, \bibinfo {author} {\bibfnamefont {M.}~\bibnamefont {Kawasaki}}, \bibinfo {author} {\bibfnamefont {Y.}~\bibnamefont {Tokura}},\ and\ \bibinfo {author} {\bibfnamefont {Y.}~\bibnamefont {Takahashi}},\ }\bibfield  {title} {\bibinfo {title} {{Magneto-optical spectroscopy on Weyl nodes for anomalous and topological Hall effects in chiral MnGe}},\ }\href {https://doi.org/10.1038/s41467-021-25276-1} {\bibfield  {journal} {\bibinfo  {journal} {Nat. Commun.}\ }\textbf {\bibinfo {volume} {12}},\ \bibinfo
  {pages} {1} (\bibinfo {year} {2021}{\natexlab{b}})}\BibitemShut {NoStop}%
\bibitem [{\citenamefont {Xu}\ \emph {et~al.}(2019)\citenamefont {Xu}, \citenamefont {Song}, \citenamefont {Wang}, \citenamefont {Weng},\ and\ \citenamefont {Dai}}]{dai19}%
  \BibitemOpen
  \bibfield  {author} {\bibinfo {author} {\bibfnamefont {Y.}~\bibnamefont {Xu}}, \bibinfo {author} {\bibfnamefont {Z.}~\bibnamefont {Song}}, \bibinfo {author} {\bibfnamefont {Z.}~\bibnamefont {Wang}}, \bibinfo {author} {\bibfnamefont {H.}~\bibnamefont {Weng}},\ and\ \bibinfo {author} {\bibfnamefont {X.}~\bibnamefont {Dai}},\ }\bibfield  {title} {\bibinfo {title} {Higher-order topology of the axion insulator ${\mathrm{euin}}_{2}{\mathrm{as}}_{2}$},\ }\href {https://doi.org/10.1103/PhysRevLett.122.256402} {\bibfield  {journal} {\bibinfo  {journal} {Phys. Rev. Lett.}\ }\textbf {\bibinfo {volume} {122}},\ \bibinfo {pages} {256402} (\bibinfo {year} {2019})}\BibitemShut {NoStop}%
\bibitem [{\citenamefont {Riberolles}\ \emph {et~al.}(2021)\citenamefont {Riberolles}, \citenamefont {Trevisan}, \citenamefont {Kuthanazhi}, \citenamefont {Heitmann}, \citenamefont {Ye}, \citenamefont {Johnston}, \citenamefont {Bud'ko}, \citenamefont {Ryan}, \citenamefont {Canfield}, \citenamefont {Kreyssig}, \citenamefont {Vishwanath}, \citenamefont {McQueeney}, \citenamefont {Wang}, \citenamefont {Orth},\ and\ \citenamefont {Ueland}}]{riberolles21}%
  \BibitemOpen
  \bibfield  {author} {\bibinfo {author} {\bibfnamefont {S.~X.~M.}\ \bibnamefont {Riberolles}}, \bibinfo {author} {\bibfnamefont {T.~V.}\ \bibnamefont {Trevisan}}, \bibinfo {author} {\bibfnamefont {B.}~\bibnamefont {Kuthanazhi}}, \bibinfo {author} {\bibfnamefont {T.~W.}\ \bibnamefont {Heitmann}}, \bibinfo {author} {\bibfnamefont {F.}~\bibnamefont {Ye}}, \bibinfo {author} {\bibfnamefont {D.~C.}\ \bibnamefont {Johnston}}, \bibinfo {author} {\bibfnamefont {S.~L.}\ \bibnamefont {Bud'ko}}, \bibinfo {author} {\bibfnamefont {D.~H.}\ \bibnamefont {Ryan}}, \bibinfo {author} {\bibfnamefont {P.~C.}\ \bibnamefont {Canfield}}, \bibinfo {author} {\bibfnamefont {A.}~\bibnamefont {Kreyssig}}, \bibinfo {author} {\bibfnamefont {A.}~\bibnamefont {Vishwanath}}, \bibinfo {author} {\bibfnamefont {R.~J.}\ \bibnamefont {McQueeney}}, \bibinfo {author} {\bibfnamefont {L.-L.}\ \bibnamefont {Wang}}, \bibinfo {author} {\bibfnamefont {P.~P.}\ \bibnamefont {Orth}},\ and\ \bibinfo {author} {\bibfnamefont {B.~G.}\ \bibnamefont {Ueland}},\
  }\bibfield  {title} {\bibinfo {title} {Magnetic crystalline-symmetry-protected axion electrodynamics and field-tunable unpinned dirac cones in euin2as2},\ }\href {https://doi.org/10.1038/s41467-021-21154-y} {\bibfield  {journal} {\bibinfo  {journal} {Nature Communications}\ }\textbf {\bibinfo {volume} {12}},\ \bibinfo {pages} {999} (\bibinfo {year} {2021})}\BibitemShut {NoStop}%
\bibitem [{\citenamefont {Yan}\ \emph {et~al.}(2022)\citenamefont {Yan}, \citenamefont {Jiang}, \citenamefont {Xiao}, \citenamefont {Lu}, \citenamefont {Song}, \citenamefont {Zhu}, \citenamefont {Luo}, \citenamefont {Sun},\ and\ \citenamefont {Yamashita}}]{yamashita22}%
  \BibitemOpen
  \bibfield  {author} {\bibinfo {author} {\bibfnamefont {J.}~\bibnamefont {Yan}}, \bibinfo {author} {\bibfnamefont {Z.~Z.}\ \bibnamefont {Jiang}}, \bibinfo {author} {\bibfnamefont {R.~C.}\ \bibnamefont {Xiao}}, \bibinfo {author} {\bibfnamefont {W.~J.}\ \bibnamefont {Lu}}, \bibinfo {author} {\bibfnamefont {W.~H.}\ \bibnamefont {Song}}, \bibinfo {author} {\bibfnamefont {X.~B.}\ \bibnamefont {Zhu}}, \bibinfo {author} {\bibfnamefont {X.}~\bibnamefont {Luo}}, \bibinfo {author} {\bibfnamefont {Y.~P.}\ \bibnamefont {Sun}},\ and\ \bibinfo {author} {\bibfnamefont {M.}~\bibnamefont {Yamashita}},\ }\bibfield  {title} {\bibinfo {title} {Field-induced topological hall effect in antiferromagnetic axion insulator candidate ${\mathrm{euin}}_{2}{\mathrm{as}}_{2}$},\ }\href {https://doi.org/10.1103/PhysRevResearch.4.013163} {\bibfield  {journal} {\bibinfo  {journal} {Phys. Rev. Res.}\ }\textbf {\bibinfo {volume} {4}},\ \bibinfo {pages} {013163} (\bibinfo {year} {2022})}\BibitemShut {NoStop}%
\bibitem [{\citenamefont {Kurumaji}\ \emph {et~al.}(2019)\citenamefont {Kurumaji}, \citenamefont {Nakajima}, \citenamefont {Hirschberger}, \citenamefont {Kikkawa}, \citenamefont {Yamasaki}, \citenamefont {Sagayama}, \citenamefont {Nakao}, \citenamefont {Taguchi}, \citenamefont {Arima},\ and\ \citenamefont {Tokura}}]{kurumaji2019skyrmion}%
  \BibitemOpen
  \bibfield  {author} {\bibinfo {author} {\bibfnamefont {T.}~\bibnamefont {Kurumaji}}, \bibinfo {author} {\bibfnamefont {T.}~\bibnamefont {Nakajima}}, \bibinfo {author} {\bibfnamefont {M.}~\bibnamefont {Hirschberger}}, \bibinfo {author} {\bibfnamefont {A.}~\bibnamefont {Kikkawa}}, \bibinfo {author} {\bibfnamefont {Y.}~\bibnamefont {Yamasaki}}, \bibinfo {author} {\bibfnamefont {H.}~\bibnamefont {Sagayama}}, \bibinfo {author} {\bibfnamefont {H.}~\bibnamefont {Nakao}}, \bibinfo {author} {\bibfnamefont {Y.}~\bibnamefont {Taguchi}}, \bibinfo {author} {\bibfnamefont {T.-h.}\ \bibnamefont {Arima}},\ and\ \bibinfo {author} {\bibfnamefont {Y.}~\bibnamefont {Tokura}},\ }\bibfield  {title} {\bibinfo {title} {Skyrmion lattice with a giant topological hall effect in a frustrated triangular-lattice magnet},\ }\href@noop {} {\bibfield  {journal} {\bibinfo  {journal} {Science}\ }\textbf {\bibinfo {volume} {365}},\ \bibinfo {pages} {914} (\bibinfo {year} {2019})}\BibitemShut {NoStop}%
\bibitem [{\citenamefont {Hirschberger}\ \emph {et~al.}(2020)\citenamefont {Hirschberger}, \citenamefont {Nakajima}, \citenamefont {Kriener}, \citenamefont {Kurumaji}, \citenamefont {Spitz}, \citenamefont {Gao}, \citenamefont {Kikkawa}, \citenamefont {Yamasaki}, \citenamefont {Sagayama}, \citenamefont {Nakao} \emph {et~al.}}]{hirschberger2020high}%
  \BibitemOpen
  \bibfield  {author} {\bibinfo {author} {\bibfnamefont {M.}~\bibnamefont {Hirschberger}}, \bibinfo {author} {\bibfnamefont {T.}~\bibnamefont {Nakajima}}, \bibinfo {author} {\bibfnamefont {M.}~\bibnamefont {Kriener}}, \bibinfo {author} {\bibfnamefont {T.}~\bibnamefont {Kurumaji}}, \bibinfo {author} {\bibfnamefont {L.}~\bibnamefont {Spitz}}, \bibinfo {author} {\bibfnamefont {S.}~\bibnamefont {Gao}}, \bibinfo {author} {\bibfnamefont {A.}~\bibnamefont {Kikkawa}}, \bibinfo {author} {\bibfnamefont {Y.}~\bibnamefont {Yamasaki}}, \bibinfo {author} {\bibfnamefont {H.}~\bibnamefont {Sagayama}}, \bibinfo {author} {\bibfnamefont {H.}~\bibnamefont {Nakao}}, \emph {et~al.},\ }\bibfield  {title} {\bibinfo {title} {High-field depinned phase and planar hall effect in the skyrmion host gd 2 pdsi 3},\ }\href@noop {} {\bibfield  {journal} {\bibinfo  {journal} {Physical Review B}\ }\textbf {\bibinfo {volume} {101}},\ \bibinfo {pages} {220401} (\bibinfo {year} {2020})}\BibitemShut {NoStop}%
\bibitem [{\citenamefont {Hirschberger}\ \emph {et~al.}(2019)\citenamefont {Hirschberger}, \citenamefont {Nakajima}, \citenamefont {Gao}, \citenamefont {Peng}, \citenamefont {Kikkawa}, \citenamefont {Kurumaji}, \citenamefont {Kriener}, \citenamefont {Yamasaki}, \citenamefont {Sagayama}, \citenamefont {Nakao} \emph {et~al.}}]{hirschberger2019skyrmion}%
  \BibitemOpen
  \bibfield  {author} {\bibinfo {author} {\bibfnamefont {M.}~\bibnamefont {Hirschberger}}, \bibinfo {author} {\bibfnamefont {T.}~\bibnamefont {Nakajima}}, \bibinfo {author} {\bibfnamefont {S.}~\bibnamefont {Gao}}, \bibinfo {author} {\bibfnamefont {L.}~\bibnamefont {Peng}}, \bibinfo {author} {\bibfnamefont {A.}~\bibnamefont {Kikkawa}}, \bibinfo {author} {\bibfnamefont {T.}~\bibnamefont {Kurumaji}}, \bibinfo {author} {\bibfnamefont {M.}~\bibnamefont {Kriener}}, \bibinfo {author} {\bibfnamefont {Y.}~\bibnamefont {Yamasaki}}, \bibinfo {author} {\bibfnamefont {H.}~\bibnamefont {Sagayama}}, \bibinfo {author} {\bibfnamefont {H.}~\bibnamefont {Nakao}}, \emph {et~al.},\ }\bibfield  {title} {\bibinfo {title} {Skyrmion phase and competing magnetic orders on a breathing kagom{\'e} lattice},\ }\href@noop {} {\bibfield  {journal} {\bibinfo  {journal} {Nature communications}\ }\textbf {\bibinfo {volume} {10}},\ \bibinfo {pages} {5831} (\bibinfo {year} {2019})}\BibitemShut {NoStop}%
\bibitem [{\citenamefont {Khanh}\ \emph {et~al.}(2020)\citenamefont {Khanh}, \citenamefont {Nakajima}, \citenamefont {Yu}, \citenamefont {Gao}, \citenamefont {Shibata}, \citenamefont {Hirschberger}, \citenamefont {Yamasaki}, \citenamefont {Sagayama}, \citenamefont {Nakao}, \citenamefont {Peng} \emph {et~al.}}]{khanh2020nanometric}%
  \BibitemOpen
  \bibfield  {author} {\bibinfo {author} {\bibfnamefont {N.~D.}\ \bibnamefont {Khanh}}, \bibinfo {author} {\bibfnamefont {T.}~\bibnamefont {Nakajima}}, \bibinfo {author} {\bibfnamefont {X.}~\bibnamefont {Yu}}, \bibinfo {author} {\bibfnamefont {S.}~\bibnamefont {Gao}}, \bibinfo {author} {\bibfnamefont {K.}~\bibnamefont {Shibata}}, \bibinfo {author} {\bibfnamefont {M.}~\bibnamefont {Hirschberger}}, \bibinfo {author} {\bibfnamefont {Y.}~\bibnamefont {Yamasaki}}, \bibinfo {author} {\bibfnamefont {H.}~\bibnamefont {Sagayama}}, \bibinfo {author} {\bibfnamefont {H.}~\bibnamefont {Nakao}}, \bibinfo {author} {\bibfnamefont {L.}~\bibnamefont {Peng}}, \emph {et~al.},\ }\bibfield  {title} {\bibinfo {title} {Nanometric square skyrmion lattice in a centrosymmetric tetragonal magnet},\ }\href@noop {} {\bibfield  {journal} {\bibinfo  {journal} {Nature Nanotechnology}\ }\textbf {\bibinfo {volume} {15}},\ \bibinfo {pages} {444} (\bibinfo {year} {2020})}\BibitemShut {NoStop}%
\bibitem [{\citenamefont {Herpin}\ and\ \citenamefont {Meriel}(1961)}]{herpin1961etude}%
  \BibitemOpen
  \bibfield  {author} {\bibinfo {author} {\bibfnamefont {A.}~\bibnamefont {Herpin}}\ and\ \bibinfo {author} {\bibfnamefont {P.}~\bibnamefont {Meriel}},\ }\bibfield  {title} {\bibinfo {title} {{\'E}tude de l'antiferromagn{\'e}tisme helicoidal de mnau2 par diffraction de neutrons},\ }\href@noop {} {\bibfield  {journal} {\bibinfo  {journal} {Journal de Physique et le Radium}\ }\textbf {\bibinfo {volume} {22}},\ \bibinfo {pages} {337} (\bibinfo {year} {1961})}\BibitemShut {NoStop}%
\bibitem [{\citenamefont {Masuda}\ \emph {et~al.}(2024)\citenamefont {Masuda}, \citenamefont {Seki}, \citenamefont {Ohe}, \citenamefont {Nii}, \citenamefont {Masuda}, \citenamefont {Takanashi},\ and\ \citenamefont {Onose}}]{masuda2024room}%
  \BibitemOpen
  \bibfield  {author} {\bibinfo {author} {\bibfnamefont {H.}~\bibnamefont {Masuda}}, \bibinfo {author} {\bibfnamefont {T.}~\bibnamefont {Seki}}, \bibinfo {author} {\bibfnamefont {J.-i.}\ \bibnamefont {Ohe}}, \bibinfo {author} {\bibfnamefont {Y.}~\bibnamefont {Nii}}, \bibinfo {author} {\bibfnamefont {H.}~\bibnamefont {Masuda}}, \bibinfo {author} {\bibfnamefont {K.}~\bibnamefont {Takanashi}},\ and\ \bibinfo {author} {\bibfnamefont {Y.}~\bibnamefont {Onose}},\ }\bibfield  {title} {\bibinfo {title} {Room temperature chirality switching and detection in a helimagnetic mnau2 thin film},\ }\href@noop {} {\bibfield  {journal} {\bibinfo  {journal} {Nature communications}\ }\textbf {\bibinfo {volume} {15}},\ \bibinfo {pages} {1999} (\bibinfo {year} {2024})}\BibitemShut {NoStop}%
\bibitem [{\citenamefont {Moon}(1982)}]{moon1982neutron}%
  \BibitemOpen
  \bibfield  {author} {\bibinfo {author} {\bibfnamefont {R.}~\bibnamefont {Moon}},\ }\bibfield  {title} {\bibinfo {title} {Neutron polarization analysis measurements on the spiral phase of mnp},\ }\href@noop {} {\bibfield  {journal} {\bibinfo  {journal} {Journal of Applied Physics}\ }\textbf {\bibinfo {volume} {53}},\ \bibinfo {pages} {1956} (\bibinfo {year} {1982})}\BibitemShut {NoStop}%
\bibitem [{\citenamefont {Matsuda}\ \emph {et~al.}(2016)\citenamefont {Matsuda}, \citenamefont {Ye}, \citenamefont {Dissanayake}, \citenamefont {Cheng}, \citenamefont {Chi}, \citenamefont {Ma}, \citenamefont {Zhou}, \citenamefont {Yan}, \citenamefont {Kasamatsu}, \citenamefont {Sugino}, \citenamefont {Kato}, \citenamefont {Matsubayashi}, \citenamefont {Okada},\ and\ \citenamefont {Uwatoko}}]{PhysRevB.93.100405}%
  \BibitemOpen
  \bibfield  {author} {\bibinfo {author} {\bibfnamefont {M.}~\bibnamefont {Matsuda}}, \bibinfo {author} {\bibfnamefont {F.}~\bibnamefont {Ye}}, \bibinfo {author} {\bibfnamefont {S.~E.}\ \bibnamefont {Dissanayake}}, \bibinfo {author} {\bibfnamefont {J.-G.}\ \bibnamefont {Cheng}}, \bibinfo {author} {\bibfnamefont {S.}~\bibnamefont {Chi}}, \bibinfo {author} {\bibfnamefont {J.}~\bibnamefont {Ma}}, \bibinfo {author} {\bibfnamefont {H.~D.}\ \bibnamefont {Zhou}}, \bibinfo {author} {\bibfnamefont {J.-Q.}\ \bibnamefont {Yan}}, \bibinfo {author} {\bibfnamefont {S.}~\bibnamefont {Kasamatsu}}, \bibinfo {author} {\bibfnamefont {O.}~\bibnamefont {Sugino}}, \bibinfo {author} {\bibfnamefont {T.}~\bibnamefont {Kato}}, \bibinfo {author} {\bibfnamefont {K.}~\bibnamefont {Matsubayashi}}, \bibinfo {author} {\bibfnamefont {T.}~\bibnamefont {Okada}},\ and\ \bibinfo {author} {\bibfnamefont {Y.}~\bibnamefont {Uwatoko}},\ }\bibfield  {title} {\bibinfo {title} {Pressure dependence of the magnetic ground states in mnp},\ }\href
  {https://doi.org/10.1103/PhysRevB.93.100405} {\bibfield  {journal} {\bibinfo  {journal} {Phys. Rev. B}\ }\textbf {\bibinfo {volume} {93}},\ \bibinfo {pages} {100405} (\bibinfo {year} {2016})}\BibitemShut {NoStop}%
\bibitem [{\citenamefont {Chen}\ \emph {et~al.}(2025)\citenamefont {Chen}, \citenamefont {Nomoto}, \citenamefont {Hirschberger},\ and\ \citenamefont {Arita}}]{chen2025}%
  \BibitemOpen
  \bibfield  {author} {\bibinfo {author} {\bibfnamefont {H.-Y.}\ \bibnamefont {Chen}}, \bibinfo {author} {\bibfnamefont {T.}~\bibnamefont {Nomoto}}, \bibinfo {author} {\bibfnamefont {M.}~\bibnamefont {Hirschberger}},\ and\ \bibinfo {author} {\bibfnamefont {R.}~\bibnamefont {Arita}},\ }\bibfield  {title} {\bibinfo {title} {Topological hall effect of skyrmions from first principles},\ }\href {https://doi.org/10.1103/PhysRevX.15.011054} {\bibfield  {journal} {\bibinfo  {journal} {Phys. Rev. X}\ }\textbf {\bibinfo {volume} {15}},\ \bibinfo {pages} {011054} (\bibinfo {year} {2025})}\BibitemShut {NoStop}%
\bibitem [{\citenamefont {Wang}\ \emph {et~al.}(2020{\natexlab{b}})\citenamefont {Wang}, \citenamefont {Morimoto},\ and\ \citenamefont {Moore}}]{moore20}%
  \BibitemOpen
  \bibfield  {author} {\bibinfo {author} {\bibfnamefont {Y.-Q.}\ \bibnamefont {Wang}}, \bibinfo {author} {\bibfnamefont {T.}~\bibnamefont {Morimoto}},\ and\ \bibinfo {author} {\bibfnamefont {J.~E.}\ \bibnamefont {Moore}},\ }\bibfield  {title} {\bibinfo {title} {Optical rotation in thin chiral/twisted materials and the gyrotropic magnetic effect},\ }\href {https://doi.org/10.1103/PhysRevB.101.174419} {\bibfield  {journal} {\bibinfo  {journal} {Phys. Rev. B}\ }\textbf {\bibinfo {volume} {101}},\ \bibinfo {pages} {174419} (\bibinfo {year} {2020}{\natexlab{b}})}\BibitemShut {NoStop}%
\bibitem [{\citenamefont {Park}\ and\ \citenamefont {Han}(2011)}]{Park2011May}%
  \BibitemOpen
  \bibfield  {author} {\bibinfo {author} {\bibfnamefont {J.-H.}\ \bibnamefont {Park}}\ and\ \bibinfo {author} {\bibfnamefont {J.~H.}\ \bibnamefont {Han}},\ }\bibfield  {title} {\bibinfo {title} {{Zero-temperature phases for chiral magnets in three dimensions}},\ }\href {https://doi.org/10.1103/PhysRevB.83.184406} {\bibfield  {journal} {\bibinfo  {journal} {Phys. Rev. B}\ }\textbf {\bibinfo {volume} {83}},\ \bibinfo {pages} {184406} (\bibinfo {year} {2011})}\BibitemShut {NoStop}%
\bibitem [{\citenamefont {Jackiw}\ and\ \citenamefont {Rebbi}(1976)}]{jackiw-rebbi}%
  \BibitemOpen
  \bibfield  {author} {\bibinfo {author} {\bibfnamefont {R.}~\bibnamefont {Jackiw}}\ and\ \bibinfo {author} {\bibfnamefont {C.}~\bibnamefont {Rebbi}},\ }\bibfield  {title} {\bibinfo {title} {Solitons with fermion number \textonehalf{}},\ }\href {https://doi.org/10.1103/PhysRevD.13.3398} {\bibfield  {journal} {\bibinfo  {journal} {Phys. Rev. D}\ }\textbf {\bibinfo {volume} {13}},\ \bibinfo {pages} {3398} (\bibinfo {year} {1976})}\BibitemShut {NoStop}%
\bibitem [{\citenamefont {Paddison}\ \emph {et~al.}(2022)\citenamefont {Paddison}, \citenamefont {Rai}, \citenamefont {May}, \citenamefont {Calder}, \citenamefont {Stone}, \citenamefont {Frontzek},\ and\ \citenamefont {Christianson}}]{paddison}%
  \BibitemOpen
  \bibfield  {author} {\bibinfo {author} {\bibfnamefont {J.~A.~M.}\ \bibnamefont {Paddison}}, \bibinfo {author} {\bibfnamefont {B.~K.}\ \bibnamefont {Rai}}, \bibinfo {author} {\bibfnamefont {A.~F.}\ \bibnamefont {May}}, \bibinfo {author} {\bibfnamefont {S.}~\bibnamefont {Calder}}, \bibinfo {author} {\bibfnamefont {M.~B.}\ \bibnamefont {Stone}}, \bibinfo {author} {\bibfnamefont {M.~D.}\ \bibnamefont {Frontzek}},\ and\ \bibinfo {author} {\bibfnamefont {A.~D.}\ \bibnamefont {Christianson}},\ }\bibfield  {title} {\bibinfo {title} {Magnetic interactions of the centrosymmetric skyrmion material ${\mathrm{gd}}_{2}{\mathrm{pdsi}}_{3}$},\ }\href {https://doi.org/10.1103/PhysRevLett.129.137202} {\bibfield  {journal} {\bibinfo  {journal} {Phys. Rev. Lett.}\ }\textbf {\bibinfo {volume} {129}},\ \bibinfo {pages} {137202} (\bibinfo {year} {2022})}\BibitemShut {NoStop}%
\end{thebibliography}%

\end{document}


\title{Supplemental Material: Gyrotropic magnetic effect in metallic chiral magnets}

\author{Nisarga Paul}
\affiliation{Department of Physics, Massachusetts Institute of Technology, Cambridge, Massachusetts 02139, USA}
\affiliation{Kavli Institute for Theoretical Physics, University of
  California, Santa Barbara, CA 93106, USA}
\author{Takamori Park}
\affiliation{Department of Physics, University of California, Santa Barbara, CA 93106, USA}
\author{Jung Hoon Han}
\affiliation{Department of Physics, Sungkyunkwan University, Suwon 16419, South Korea}
\author{Leon Balents}
\affiliation{Kavli Institute for Theoretical Physics, University of
  California, Santa Barbara, CA 93106, USA}
\affiliation{French American Center for Theoretical Science, CNRS,
  KITP, Santa Barbara, CA 93106-4030, USA}
\affiliation{Canadian Institute for Advanced Research, Toronto, Ontario, M5G 1M1, Canada}

\date{\today}

\maketitle

\tableofcontents

\section{Theory of optical rotation} 
\label{app:opticalrotation}

We review the theory of optical rotation. The permittivity tensor $\varepsilon(\omega,\vec k)$ relates the displacement field $\vec D$ and electric field $\vec E$,
\begin{equation}
    D_i = \varepsilon_{ij}(\omega,\vec k) E_j,
\end{equation}
and satisfies $\varepsilon_{ij}(-\omega,-\vec k) = \varepsilon_{ji}^*(\omega,\vec k)$. Expanding its antisymmetric part $\varepsilon^A_{ij} = (\varepsilon_{ij}-\varepsilon_{ji})/2$ at small $\vec k$ gives
\begin{equation}
    \varepsilon^A_{ij}(\omega,\vec k) = \varepsilon^A_{ij}(\omega) + i\gamma_{ijl}^A(\omega) k_l + \cdots 
\end{equation}
where $\gamma_{ijl}^A = -\gamma_{jil}^A$. This tensor describes natural gyrotropy effects including natural optical
rotation and natural circular dichroism. It is instructive to introduce the gyration (pseudo)tensor $g_{ij}$ according to 
\begin{align}
    \gamma_{ijl}^A &= (c/\omega)\epsilon_{ijm} g_{ml}\\
    g_{ij} &= (\omega/2c)\epsilon_{ilm}\gamma^A_{lmj}.
\end{align}
The gyration tensor is related to the tensor $\agme_{ij}$ via~\cite{souza16} (note tranposed indices)
\begin{equation}\label{eq:gij}
    g_{ij} = \frac{1}{\omega c \epsilon_0} \pa{\agme_{ji} - \bar \alpha^{\text{GME}} \delta_{ij}}
\end{equation}
where $\bar \alpha^{\text{GME}} = \Re(\Tr\agme)$. A medium with $g_{ij}\neq 0$ is called optically active. In an optically active medium, left- and right-handed circularly polarized light have different velocities and refractive indices. When plane-polarized light passes through such a medium, the plane of polarization is rotated by an angle $\phi$ proportional to the thickness $d$ of the material. This is captured by the optical activity coefficient, or \textit{rotatory power}
\begin{equation}
    \rho = \phi/d.
\end{equation}
For example, $|\rho| \approx 0.328$ rad/mm in quartz for light of wavelength $\lambda = 0.63 \mu m$ \cite{Newnham2005}. In general, for light which propagates along the unit vector $\hat{\vec n}$, we have 
\begin{equation}
    \rho = \frac{\pi}{\lambda} g_{ij}\hat n_i\hat n_j,
\end{equation}
assuming birefringence is small (i.e. $\epsilon_{ij}(0,0)\approx \epsilon_0 \delta_{ij}$). Together with Eq. \eqref{eq:gij}, we have
\begin{subequations}
    \begin{align}
        \rho &= \frac{\pi}{\lambda \omega c \varepsilon_0 } \pa{\agme_{ij} - \bar \alpha^{\text{GME}} \delta_{ij}}\hat n_i \hat n_j\\
        &= \frac{1}{2c^2\varepsilon_0}\pa{ \agme_{ij}\hat n_i \hat n_j - \bar\alpha^{\text{GME}}}    \end{align}
\end{subequations}
While only the symmetric part of $g_{ij}$ enters optical activity, the antisymmetric part need not vanish; it gives rise to a ``transverse" GME which is associated to an overall electric-field-induced magnetization. In polar metals \cite{Agranovich,souza16} $g_{ij}$ will generally have an antisymmetric part, with an associated polar vector $\delta_i = \epsilon_{ijk} g_{jk}$. \par 
For any crystal with cubic symmetry or higher, $\agme_{ij} = \bar\alpha^{\text{GME}}\delta_{ij}/3$ and the above reduces to 
\begin{equation}
    \rho = -\frac{\bar\alpha^{\text{GME}}}{3c^2\varepsilon_0}. 
\end{equation}
The gyration tensor is constrained by point group symmetry. First of all, it vanishes in the presence of an inversion symmetry. If there is no inversion symmetry and the $z$ axis is normal to a plane with rotational symmetry group $S$, the only nonvanishing elements of the (symmetrized) gyration tensor are listed in Table \ref{tab:g-tensor}. 

\begin{table}[t]
\centering
\caption{Nonvanishing $g$-tensor components under different symmetry groups \cite{Landau1984}.}
\label{tab:g-tensor}
\begin{tabular}{ll}
\hline
\textbf{Symmetry, $S$} & \textbf{$g$-tensor components} \\
\hline
$C_1$ 
  & all components \\
$C_2$ 
  & $g_{xx}, g_{yy}, g_{zz}, g_{xy}$ (the last can be made zero by rotation in $xy$) \\
$C_s$ 
  & $g_{xy}, g_{yz}$ (one can be made zero by $xy$ rotation) \\
$C_{2v}$ 
  & $g_{xy}$ only (mirror planes: $xz$, $yz$) \\
$D_2$ 
  & $g_{xx}, g_{yy}, g_{zz}$ \\
$C_3, C_4, C_6, D_3, D_4, D_6$ 
  & $g_{xx} = g_{yy}$, $g_{zz}$ \\
$S_4$ 
  & $g_{xx} = -g_{yy}, g_{xy}$ (one can be made zero by $xy$ rotation) \\
$D_{2d}$ 
  & $g_{xy}$ (where $x$, $y$ axes lie in vertical symmetry planes) \\
$T, O$  
  & $g_{xx} = g_{yy} = g_{zz}$ \\
\hline
\end{tabular}
\end{table}
\section{Distinguishing the GME and the Faraday effect}
\label{sec:GMEvsFaraday}

Optical activity is distinct from the Faraday effect, which also describes rotation of plane-polarized light but occurs due to broken $\mathcal{T}$ and does not require $\vec q$-dependence of $\varepsilon_{ij}$. Let us briefly recap the Faraday effect. It is useful to write $\varepsilon_{ij} = \varepsilon_{ij}'+i\varepsilon_{ij}''$ and $\epsilon_{ij}^{-1} = \eta_{ij} = \eta_{ij}' + i\eta_{ij}''$. The axial vector dual to the antisymmetric form $\eta_{ij}''$ is defined as $G_i = \frac12 \epsilon_{ijk} \eta_{jk}''$ (where $\varepsilon_{ijk}$ is the Levi-Civita symbol) and the gyration vector $g_j$ is defined via $G_i = -\frac{1}{|\varepsilon|} \varepsilon_{ij}'g_j$, where $|\varepsilon| = \det \varepsilon_{ij}$. The Faraday rotation per unit length is then given by $\tilde\Theta_F/d = (\omega g/2cn_0)\cos\theta$, where $n_0$ is the refractive index and $\theta$ is the angle between the direction of light propagation $\hat{\vec n}$ and the gyration vector~\cite{Landau1984}. In thin films, the Faraday rotation is simply related to the optical conductivity as~\cite{Han2022May,shimano2011terahertz}
\begin{equation}
    \tilde \Theta_F(\omega)/d = \sigma_{xy}(\omega)/(d\sigma_{xx}(\omega) + (n_0+1)/Z_0)
\end{equation}
where $Z_0$ is the vacuum impedance and it is assumed that $\hat{\vec n}=\hat z$. Hence the low-frequency limit of Faraday rotation is simply related to the Berry curvature of Bloch bands in the ultraclean limit where the effects of impurities can be ignored, and more generally Faraday rotation is related to the optical Hall conductivity. In contrast, the low-frequency limit of optical activity probes the magnetic moments of the Bloch bands. The related effect corresponding to a rotation of plane-polarized light upon \textit{reflection} due to broken $\mathcal{T}$ is the Kerr effect. \par 

Optical activity and the Faraday effect can be distinguished by measuring optical rotation angles $\theta_{\hat{\vec n}}$ and $\theta_{-\hat{\vec n}}$ along two opposite directions $\hat{\vec n}$ and $-\hat{\vec n}$. Optical activity corresponds to the sum $(\theta_{\hat{\vec n}} + \theta_{-\hat{\vec n}})/2$ while the Faraday effect corresponds to the difference $(\theta_{\hat{\vec n}} - \theta_{-\hat{\vec n}})/2$. 

\par

\begin{table}
\centering
\begin{tabular}{|l|c|c|}
\hline
 & optical activity / GME & Faraday effect \\
\hline
spin cycloid & \texttimes & \texttimes \\
\hline
spin spiral & \checkmark & \texttimes \\
\hline
multi-$q$ texture & \checkmark & \checkmark \\
\hline
\end{tabular}
\caption{\textbf{GME vs. Faraday. } Spin textures considered in the main text, and whether or not electrons Hunds coupled to these spin textures can give rise to a finite optical activity/ GME or a finite Faraday rotation.}\label{tab:GMEvsFaraday}
\end{table}

While broken $\mathcal{T}$ is necessary for the Faraday effect, it is not sufficient; there could exist further symmetries which enforce a vanishing Faraday effect. This is the case for the spin cycloid and spin spiral textures with vanishing magnetization coupled to either nonrelativistic electrons or (a pair of $\mathcal{T}$-related) Weyl nodes. One way to see this is to consider the antiunitary symmetry $A = T_{a/2}\mathcal{T}$, where $T_{a/2}$ is a half-period translation along the wavevector of the cycloid or spiral and $\mathcal{T} = i\sigma^y K$ is physical time-reversal. $A$ preserves the kinetic terms $H_0^{\text{NR}} = p^2/2m$, the nonrelativistic case, and $H_0^{\text{R}} = \sum_w v_f(\vec p\cdot \vec \sigma)_w$, a pair of $\mathcal{T}$-related Weyl nodes. $A$ also preserves the Hund's coupling term $H'=J\sum_{\vec r} \bm S_{\vec r}\cdot\vec{\sigma}$ when $\bm S_{\vec r} = (\cos \vec q \cdot \vec r,0,\sin \vec q \cdot \vec r)$ or $(\cos \vec q \cdot \vec r,\sin \vec q \cdot \vec r, 0)$. Due to the symmetry $A$, the Berry curvature of any Bloch band must satisfy $\Omega(\vec k) = -\Omega(-\vec k)$ and the Hall conductivity must vanish; indeed, the optical Hall conductivity must vanish at all frequencies. This implies that the Faraday rotation vanishes in these cases. (In fact, for the nonrelativistic case, a stronger statement holds; upon a spin-rotation, we can arrange that $S^y = 0$, so the Hamiltonian is invariant under complex conjugation $K$, while $\Omega(\vec k)$ changes sign, and hence the Berry curvature vanishes pointwise). In contrast, for the multi$-q$ spin texture, the Faraday effect can in general be nonvanishing. We summarize these statements in Table \ref{tab:GMEvsFaraday}. Spin spirals and cycloids (i.e. Bloch spirals and N\'eel spirals) are common ground states in chiral magnets even in the absence of a magnetic field, arising for instance due to the Dzyaloshinskii–Moriya interaction, RKKY interaction, or frustration. The GME therefore provides a unique transmission-based optical probe for this broad and physically important class of magnetic ground states in chiral systems. 

\par 
In summary, there are two transmission-type effects which cause the rotation of plane-polarized light: optical activity (whose low-frequency limit is the GME) and the Faraday effect. Optical activity requires broken $\mathcal{P}$, while the Faraday effect requires broken $\mathcal{T}$. However, broken $\mathcal{T}$ is not sufficient for the Faraday effect, as in the case of electrons coupled to a spin cycloid or spin spiral texture, as summarized by Table \ref{tab:GMEvsFaraday}. Finally, the contributions of optical activity and the Faraday effect to optical rotation can be easily separated by measuring along opposite directions and taking the sum and difference. \par

\section{Details of perturbation theory}
\label{sec:perturbation}
Here, we derive the leading contributions to the GME response coefficient to leading order in small $J$. We will denote as $A^{(n)}$ the $O(J^n)$ term in the expansion of any quantity $A$. We work in the clean limit $\tau\to \infty$. The leading contributions to $\alpha_{ij}^\textrm{GME}$ take the form 
\begin{subequations}\label{eqs:alphaGME012}
    \begin{align}
        \alpha_{ij}^{\text{GME},(0)} &= 0\\
        \alpha_{ij}^{\text{GME},(1)} &= -\frac{e}{\hbar}\sum_n\int[d\vec k] f'(\varepsilon_{\vec kn}^{(0)})\left(\hbar v_{\vec kn,i}^{(0)} m_{\vec k n,j}^{(1)} - \varepsilon_{\vec kn}^{(1)}\partial_{i} m_{\vec kn,j}^{(0)}\right)\\
        \alpha_{ij}^{\text{GME},(2)} &= -\frac{e}{\hbar} \sum_n \int [d\vec k] f'(\varepsilon_{\vec kn}^{(0)})
        \left(\hbar v_{\vec kn,i}^{(0)} m_{\vec kn,j}^{(2)} - \varepsilon_{\vec kn}^{(1)} \partial_i m_{\vec kn,j}^{(1)} - \varepsilon_{\vec kn}^{(2)} \partial_i m_{\vec kn,j}^{(0)}\right) -\frac12 f''(\varepsilon_{\vec kn}^{(0)})(\varepsilon_{\vec kn}^{(1)})^2 \partial_i m_{\vec kn,j}^{(0)}
    \end{align}
\end{subequations}
where $\partial_i = \partial_{k_i}$ and $\alpha_{ij}^{\text{GME},(0)}$ vanishes due to the presence of $\mathcal{P}$ symmetry when $J=0$. We split up the discussion for the relativistic and non-relativistic cases.

\textbf{Relativistic case. } In the relativistic case, the unperturbed Hamiltonian and the perturbation are given by $H_0^{\text{R}}=\sum_w\chi_w v_f(\vec p \cdot\vec\sigma)_w$ and
$H'=J\bm S_{\vec r}\cdot\bm \sigma$, respectively. In order to calculate the perturbative corrections to the gyrotropic response, we first calculate the perturbative corrections to the energies and the eigenstates. We reformulate standard nondegenerate perturbation theory using projection operators of states $\rho_{\q,\alpha}\equiv\ketbra{\psi_{\q,\alpha}}{\psi_{\q,\alpha}}$ instead of the states $\ket{\psi_{\q,\alpha}}$. The reason for this is simply that the projection operator for the eigenstate of $H_0^\text{R}$ has a simpler representation then the eigenstate itself, and no information is lost by using the projection operator.

First, we present the the general formulas for nondegenerate perturbation theory using projection operators. Let $\{\varepsilon_n^{(0)}\}$ and $\{\rho^{(0)}_n\}$ denote the complete set of eigenvalues and eigenstate projection operators of the unperturbed Hamiltonian. Then, it's straightforward to see that the standard nondegenerate perturbation theory formulas for energy correction can be expressed as
\begin{subequations}
\begin{align}
    \varepsilon_n^{(1)}=&\tr[\rho^{(0)}_n H'],\\
    \varepsilon_n^{(2)}=&\sum_{m\neq n}\frac{\tr[\rho^{(0)}_nH'\rho_m^{(0)}H']}
    {\varepsilon_n^{(0)}-\varepsilon_m^{(0)}}.
\end{align}
\end{subequations}
The corrections to the wavefunctions becomes corrections to the eigenstate projection operators.
\begin{subequations}
    \begin{align}
        \rho_n^{(1)}=&\sum_{m\neq n}\frac{\rho^{(0)}_m H'\rho_n^{(0)}}
        {\varepsilon_n^{(0)}- \varepsilon_m^{(0)}}+\textrm{H.c.}\\
        \rho_n^{(2)}=&\frac{1}{2}\sum_{m,l\neq n}
        \frac{\rho^{(0)}_mH'\rho^{(0)}_nH'\rho^{(0)}_l}
        {(\varepsilon_n^{(0)}-\varepsilon_m^{(0)})(\varepsilon_n^{(0)}-\varepsilon_l^{(0)})}
        +\sum_{m,l\neq n}\frac{\rho^{(0)}_mH'\rho^{(0)}_lH'\rho^{(0)}_n}
        {(\varepsilon_n^{(0)}-\varepsilon_m^{(0)})(\varepsilon_n^{(0)}-\varepsilon_l^{(0)})}\notag\\
        &-\frac{1}{2}\sum_{m\neq n}
        \frac{\rho_m^{(0)}H' \rho_n^{(0)} H'\rho_n^{(0)}}{(\varepsilon_n^{(0)}-\varepsilon_m^{(0)})^2}
        -\frac{1}{2}\rho_n^{(0)}
        \sum_{m\neq n}\frac{\tr[\rho_m^{(0)}H'\rho_n^{(0)}H']}{(\varepsilon_n^{(0)}-\varepsilon_m^{(0)})^2}
        +\textrm{H.c.}
    \end{align}
\end{subequations}
It's easy to verify that both corrections are traceless, $\tr[\rho_n^{(1)}]=\tr[\rho_n^{(2)}]=0$.
These formulas can now be applied to calculate the corrections to the case when the unperturbed Hamiltonian is given by $H_0^\textrm{R}$ and $H'=J\sum_{\vec r} \bm S_{\vec r}\cdot\vec{\sigma}$.

\par 
The energies and projection operators of the eigenstates of $H_0^\textrm{R}$ are given by $\varepsilon_{\vec q,\alpha}^{(0)}=\alpha\chi_w v_F\abs{\vec q}$ and $\rho_{\vec q,\alpha}^{(0)}=\frac{1}{2}(I+\alpha\hat{\vec q}\cdot\vec{\sigma})\otimes \ketbra{\vec q}{\vec q}$, respectively, where $\alpha=\pm$ labels the bands and $\ket{\vec q}$ is a momentum eigenstate. In this notation, the perturbation is expressed as $H'=J\sum_{\vec G}\bm{S}_{\vec G}\cdot\vec{\sigma}\otimes\sum_{\vec q}\ketbra{\vec q+\vec G}{\vec q}$. For convenience, we define $\tilde\rho_{\vec q,\alpha}^{(0)}\equiv\frac{1}{2}(I+\alpha\hat{\vec q}\cdot\vec{\sigma})$. The corrections to the energies can then be written as
\begin{subequations}
\begin{align}\label{eq:energycorrections}
    \varepsilon_{\vec q,\alpha}^{(1)}=&J\tr[\tilde \rho^{(0)}_{\vec q,\alpha}\qty(\bm{S}_{\vec0}\cdot\vec\sigma)]\\
    \varepsilon_{\vec q,\alpha}^{(2)}=&J^2\sum_{\beta=\pm}\sum_{\vec G\neq 0}
\frac{\tr\qty[\tilde{\rho}^{(0)}_{\vec q,\alpha}
  \bm{S}_{-\vec G}\cdot\vec{\sigma}\tilde{\rho}^{(0)}_{\vec q+\vec G,\beta}
  \bm{S}_{\vec G}\cdot\vec{\sigma}]}
  {\varepsilon^{(0)}_{\vec q,\alpha}-\varepsilon^{(0)}_{\vec q +\vec G,\beta}}
\end{align}
\end{subequations}
and the first order correction to the eigenstate projection operator is 
\begin{align}\label{eq:rho1qalpha}
\rho^{(1)}_{\vec q,\alpha}=&
J\sum_{\vec G\neq0}\sum_{\beta=\pm}\frac{\trhozero{q+G}{\beta}\,
\bm S_\vec{G}\cdot\vec\sigma\,\trhozero{q}{\alpha}}{\epzero{q}{\alpha}-\epzero{q+G}{\beta}}
\otimes\ket{\q+\G}\bra{\q}+J\frac{\trhozero{q}{\alpha}\,\bm{S}_\vec{0}\cdot\vec\sigma\,\trhozero{q}{-\alpha}}
{\epzero{q}{\alpha}-\epzero{q}{-\alpha}}\otimes\ket{\q}\bra{\q}
+\textrm{H.c.}.
\end{align}
The second order correction contains many terms, but luckily we only need the projector in the $\q$ momentum sector, 
$\ev**{\rho^{(2)}_{\vec q,\alpha}}{\q}$. We find that this is
\begin{align}\label{eq:rho2qalpha}
  \ev**{\rho^{(2)}_{\vec q,\alpha}}{\q}=&J^2\sum_{\substack{\G\neq0\\ \beta=\pm}}
  \frac{\trhozero{q}{-\alpha}\,\bm{S}_{-\G}\cdot\vec\sigma\,\trhozero{q+G}{\beta}\,\bm{S}_\G\cdot\vec\sigma\,
  \trhozero{q}{\alpha}+\textrm{h.c.}}{\qty(\epzero{q}{\alpha}-\epzero{q+G}{\beta})\qty(\epzero{q}{\alpha}-\epzero{q}{-\alpha})}\notag-J^2\sum_{\substack{\vec G\neq0\\ \beta=\pm}}\frac{\tr[\trhozero{q}{\alpha}\,\bm{S}_{-\G}\cdot\vec\sigma\,\trhozero{q+G}{\beta}\,\bm{S}_\G\cdot\vec\sigma]}
  {\qty(\epzero{q}{\alpha}-\epzero{q+G}{\beta})^2}\trhozero{q}{\alpha}\notag\\
  &-J^2\frac{\trhozero{q}{\alpha}\,\bm{S}_{\vec0}\cdot\vec\sigma\,\trhozero{q}{\alpha}\,\bm{S}_{\vec0}\cdot\vec\sigma\,
  \trhozero{q}{-\alpha}+\textrm{h.c.}}{\qty(\epzero{q}{\alpha}-\epzero{q}{-\alpha})^2}\notag+J^2\frac{\trhozero{q}{-\alpha}\,\bm{S}_{\vec0}\cdot\vec\sigma\,\trhozero{q}{\alpha}\,\bm{S}_{\vec0}\cdot\vec\sigma\,
  \trhozero{q}{-\alpha}}{\qty(\epzero{q}{\alpha}-\epzero{q}{-\alpha})^2}\notag-J^2\frac{\tr[\trhozero{q}{\alpha}\,\bm{S}_{\vec0}\cdot\vec\sigma\,\trhozero{q}{-\alpha}\,\bm{S}_{\vec0}\cdot\vec\sigma]}
  {\qty(\epzero{q}{\alpha}-\epzero{q}{-\alpha})^2}\trhozero{q}{\alpha}.
\end{align}
Now that we have the perturbative corrections to the energy and eigenstate projection operator, we can calculate the orbital and spin magnetizations.
Before that, we need to fold the bands into the first Brillouin zone. Since we are only interested in the two bands closest to the Weyl node, we simply assume $\q$ is in the first Brillouin zone and treat $\alpha=\pm$ as a band index for these two bands.
Next, we define $e^{-i\k\cdot\r}\rho_{\k\alpha}e^{i\k\cdot\r}\equiv P_{\k \alpha}=\ketbra{u_{\k\alpha}}{u_{\k\alpha}}$ as the projection operator of the periodic part of the eigenstate. The two are related by $\mel{\q}{P_{\k\alpha}}{\q'}
=\mel{\k+\q}{\rho_{\k\alpha}}{\k+\q'}$.
The orbital and spin magnetization can then be expressed using this projection operator as
\begin{align}
    m_{\k\alpha,a}^\textrm{orb}=&-\frac{ie}{2\hbar}\epsilon_{abc}\Tr[\partial_b  P_{\k\alpha}\,H_\k\partial_c\,P_{\k\alpha}]\\
    m_{\k\alpha,a}^\textrm{spin}=&-\frac{e\hbar g_s}{4m_e}\Tr[\sigma_a P_{\vec k\alpha}]
\end{align}
where $H_\k=e^{-i\k\cdot\r}He^{i\k\cdot\r}$ is the Bloch Hamiltonian. From this formula, we obtain the first and second order corrections to magnetization $\vec m_{\k\alpha}^{(1)},\vec m_{\k\alpha}^{(2)}$. The velocity can be obtained as $v_{\alpha i}^{(n)}=\hbar^{-1}\partial_i \varepsilon_{\alpha}^{(n)}$. We note that several simplifications occur. Firstly, the second-order corrections come from the pairs $\bm{S}_\G,\bm{S}_{-\G}$, so we calculate the contribution from each of these pairs separately and then take their sum. In addition, summing over $\chi_w=\pm1$ before integrating over the Fermi surface simplifies the integrand. Following these steps, we obtained the formulas in the main text, which we also reproduce below.\par 

\textbf{Non-relativistic case. } Next we discuss the non-relativistic case, which has unperturbed Hamiltonian $H_0^{\text{NR}} = p^2/2m$ and doubly degenerate unperturbed energies, $\varepsilon_{\vec q,\alpha}^{(0)} \equiv \varepsilon_{\vec q}^{(0)}=q^2/2m$. We consider the eigenstate projection operator $\rho_{\vec q} = \sum_{\alpha=\pm} \rho_{\vec q,\alpha}$ onto the two lowest eigenstates, with $\rho_{\vec q}^{(0)} = \ket{\vec q}\bra{\vec q}$. We have 
\begin{subequations}
\begin{align}
        \varepsilon_{\vec q,\alpha}^{(1)} &= \alpha J |\bm S_{0}|, \\
        \rho_{\vec q}^{(1)} &= J\sum_{\vec G\neq 0} \frac{\bm S_{\vec G}\cdot \bm \sigma}{\varepsilon_{\vec q}^{(0)} - \varepsilon_{\vec q+\vec G}^{(0)}}\ket{\vec q+\vec G}\bra{\vec q} + \text{H.c.}.
\end{align}
\end{subequations}
Several simplifications occur. First, because $P_{\vec q}^{(0)} \equiv e^{-i\vec q\cdot \vec r} \rho_{\vec k}^{(0)} e^{i\vec q\cdot \vec r}$ is $\vec q$-independent, $\vec m^{\text{orb},(0)}_{\alpha}$ and $\vec m^{\text{orb},(1)}_{\alpha}$ vanish, and some algebra shows that $\vec m^{\text{orb},(2)}_{\alpha}$ vanishes as well. Therefore $\alpha_{ij}^{\text{GME,NR,orb}} = 0$
to at least $O(J^2)$, so we only concern ourselves with $\alpha_{ij}^{\text{GME,NR,spin}}$. Next, because $\Tr[\bm \sigma P_{\vec k\alpha}^{(0)}] = \alpha \hat{\bm S}_0$ and $\Tr[\bm \sigma P_{\vec k\alpha}^{(1)}] = 0$, the leading GME contribution is simply
\begin{equation}
    \alpha^{\text{GME,NR,spin}(2)}_{ij} = -e\int[d\vec k] f'(\varepsilon_{\vec k }^{(0)}) v_{\vec k,i}^{(0)}\sum_{\alpha} m_{\vec k\alpha,j}^{\text{spin}(2)} = \frac{e^2\hbar g_s}{4m_e} \int [d\vec k] f'(\varepsilon_{\vec k}^{(0)}) v_{\vec k,i}^{(0)} \Tr[\sigma_j P_{\vec k}^{(2)}].
\end{equation}
From second order perturbation theory, we find
\begin{equation}
\Tr[\sigma_j P_{\vec k}^{(2)}] = \sum_{\vec G\neq \vec 0} \frac{4i(\bm S_{\vec G}\times \bm S_{-\vec G})_j}{(\varepsilon_{\vec k}^{(0)} - \varepsilon_{\vec k+\vec G}^{(0)})^2}
\end{equation}
and at sufficiently low temperatures, $f'(\varepsilon_{\vec k}^{(0)}) =-\delta(|k|-k_F)m/\hbar k_F$, from which we obtained the result in the main text, which we also reproduce below.\par 

\textbf{Final results. } Here we reproduce the perturbative results for the GME tensors: 

\begin{subequations}
\begin{align}
\alpha_{ij}^{\textrm{GME,NR,spin}} &= \frac{i\omega \tau}{1-i\omega\tau}  \alpha_0 \frac{g_s J}{4 m_e (v_F^{\rm NR} )^2} \sum_{\vec G\neq 0} \,  f_{\vec G , ij} g^{\text{NR}}(\xi) \,\,    +\cdots\quad   \\
\alpha_{ij}^{\textrm{GME,R,orb}}&=
  \frac{i\omega \tau}{1-i\omega\tau}  \alpha_0 \sum_{\vec G\neq0}\frac{J}{\hbar v_F G} 
  f_{\vec G} \qty[g_{\perp}^\textrm{R,orb}\qty(\xi) P^\perp_{\vec G , ij}
+g_{\|}^\textrm{R,orb}\qty(\xi)  P^\parallel_{\vec G , ij} ] +\cdots  \\
\alpha_{ij}^{\textrm{GME,R,spin}} &= \frac{i\omega \tau}{1-i\omega\tau}  \alpha_0 \frac{g_s J}{4 m_e v_F^2}
  \sum_{\vec G\neq0} \bigg( f_{\vec G}\qty[g_{\perp}^\textrm{R,spin}\qty(\xi) P^\perp_{\vec G , ij}
+g_{\|}^\textrm{R,spin}\qty(\xi) P^\parallel_{\vec G , ij} ] +  f'_{\vec G , ij} g_L\qty(\xi) \bigg)  +\cdots
\end{align}
\label{eq:perturb}
\end{subequations}
The prefactor $\alpha_0  = \frac{e^2}{h^2}J$ carries the dimension of the GME tensor. We introduced the non-relativistic velocity $v_F^{\rm NR} = \hbar k_F /m$ in terms of the Fermi momentum $k_F$ (Fermi energy $E_F = \hbar^2 k_F^2 /2m$) and the renormalized electron mass $m$. The Fermi momentum $k_F$ is related to the Fermi energy $E_F$ in the relativistic case as $E_F = \hbar v_F k_F$. In deriving the above formulas, the dimensionless quantity $\xi = 2k_F/G$ ($G=|{\bf G}|$) is assumed to satisfy $|\xi|<1$ to rule out scattering among the same-energy states and the opening of the minigap. Terms in $\cdots$ are subleading in $J/E_K^{\text{NR}}, J/E_K^{\text{R}},$ and $J/m_ev_F^2$, where $E_K^{\text{NR}} = \hbar^2 G^2/2m$ or $E_K^{\text{R}} = \hbar v_F G$ is the characteristic electronic kinetic energy. We refer to these quantities as $E_K$ when the context is clear. The symmetric part of $\alpha_{ij}$ contributes to the optical activity, while the antisymmetric part contributes to a ``transverse GME" effect~\cite{souza16}. 
\par 
Form factors $f_{\vec G , ij } = i \vec{\hat{G}}_i (\bm S_{\vec G} \times \bm S_{-\vec G})_j $ and $f_{\bf G} = \Tr [f_{\vec G,ij} ]$ are introduced in Eq.~\eqref{eq:perturb} in terms of $\vec{\hat{G}}=\vec{G}/G$, where $G = 2\pi /\lambda$ is related to the spin texture wavelength $\lambda$. Transverse and longitudinal projectors are defined by $P^\perp_{\vec G , ij} = \delta_{ij} - \vec{\hat{G}}_i\vec{\hat{G}}_j$ and $P^\parallel_{\vec G, ij} = \vec{\hat{G}}_i \vec{\hat{G}}_j$. The form factor $f'_{\vec G , ij} \equiv f_{\vec G , ij} - f_{\vec G} P^\parallel_{\vec G , ij}$ appearing in the last equation is equivalent to $i \vec{\hat G}_i \qty[{\bm S}_{\vec G} \times {\bm S}_{-\vec G} ]^\perp_j$ where $\qty[ {\bm S}_{\vec G} \times {\bm S}_{-\vec G} ]^\perp = {\bm S}_{\vec G} \times {\bm S}_{-\vec G} -  \vec{\hat G}\cdot \qty( {\bm S}_{\vec G} \times {\bm S}_{-\vec G} ) \vec{\hat G}$ refers to the transverse component. The form factor $f_{\vec G}$ shows up only in spin spirals rotating in the plane {\it perpendicular} to the propagation vector $\vec{G}$. Spin cycloids with spins rotating in the plane containing ${\bf G}$ may only contribute to $\alpha_{ij}^{\textrm{GME,R,spin}}$. The non-relativistic orbital magnetic moment and $\alpha_{ij}^{\text{GME,NR,orb}}$ are highly suppressed, presumably due to the absence of spin-orbit coupling and the lack of Berry curvature in the unperturbed bands. \par 

The scaling functions are given by
\begin{subequations}
\begin{align}
  g^{\text{NR}}(\xi) =& 2\xi^2 \left(\frac{\xi}{\xi^2-1} +\tanh^{-1}\xi \right)\\
  g_{\perp}^\textrm{R,orb}(\xi)=&2\qty(1-\xi^2)\qty(\xi-\tanh^{-1}\xi)/\xi^3\\
  g_{\|}^\textrm{R,orb}(\xi)=& 2\frac{\xi(\xi^4-4\xi^2+2)-(3\xi^4-5\xi^2+2)\tanh^{-1}\xi}{\xi^3(\xi^2-1)}\\
  g_{\perp}^\textrm{R,spin}(\xi)=&(\xi-\xi^3-\tanh^{-1}\xi)/\xi^2\\
  g_{\|}^\textrm{R,spin}(\xi)=&\frac{(2-3\xi^2)(\xi+(\xi^2-1)\tanh^{-1}\xi)}{\xi^2(\xi^2-1)}\\
  g_L(\xi)=&-\xi+\tanh^{-1}\xi
\end{align}
   \label{eq:scalingfns}
\end{subequations}
\noindent which are plotted in Fig. 2a of the main text. In general, the orbital part of the 
scaling functions are even in $\xi$ and finite at $\xi = 0$. The spin parts are odd in $\xi$ and vanish when $\xi = 0$. Only the $\xi \ge 0$ part is relevant for the non-relativistic scaling function $g^{\text{NR}}(\xi)$, whereas both signs of $\xi$ are meaningful for relativistic electrons since the Fermi energy can be of both signs. Spin contributions vanish for the Weyl electrons at charge neutrality ($E_F = 0$) while for non-relativistic electrons they remain finite since $\xi > 0$. The high-density region ($\xi > 1$) not captured by the scaling formulas can be treated by solving the Hamiltonian numerically.

\section{Details of numerical approach}
\label{app:alphanumerical}

Here we summarize the numerical calculations, starting with the numerical solution to the bandstructure. We are solving 
\begin{equation}\label{eq:A1}
H = H_0(\vec p) + J\sum_{\vec r} \bm S_{\vec r} \cdot \bm\sigma
\end{equation}
for two cases, $ H_0^{\text{NR}} = p^2/2m$ and $H_0^{\text{R}} = \sum_{w} \chi_w v_F (\vec p \cdot \bm \sigma )_w$. For the relativistic case, we will suppress the sum over Weyl nodes. Final quantities are summed over two Weyl nodes of opposite chirality. We assume $\bm S_{\vec r + \vec a_i} = \bm S_{\vec r}$ for some primitive vectors $\vec a_1,\vec a_2,\vec a_3$ which generate a lattice $\Gamma$ in the case of the multi-spiral textures and a single vector $\vec a_1$ in the case of the single spiral or cycloid. Let $\vec G_i$ satisfy $\vec G_i \cdot \vec a_j = 2\pi \delta_{ij}$, and let $\Lambda$ be the reciprocal lattice generated by $\vec G_i$. Let $\vec k$ belong to the first Brillouin zone (BZ) of $\Lambda$. By Bloch's theorem, a solution to $H\psi = E\psi$ can be expressed
\begin{equation}
    \psi(\vec r) = e^{i\vec k \cdot \vec r} u_{\vec k}(\vec r)
\end{equation}
where $H_{\vec k}u_{\vec k} = E u_{\vec k}$ and $H_{\vec k}  = e^{-i\vec k \cdot \vec r} H e^{i\vec k \cdot \vec r}$ is the Bloch Hamiltonian at $\vec k$. We let 
\begin{equation}
        u_{\vec k}(\vec r) = \sum_{\vec G \in \Lambda}  u_{\vec k,\vec G} e^{i \vec G \cdot \vec r},\qquad \bm S_{\vec r} = \sum_{\vec G\in \Lambda} \bm S_{\vec G} e^{i\vec G\cdot \vec r}.
\end{equation}
For the non-relativistic and relativistic cases, respectively, we have
\begin{subequations}
    \begin{align}
    H^{\text{NR}}_{\vec k} u_{\vec k} &= \sum_{\vec G,\vec G'} (\hbar^2 \delta_{\vec G,\vec G'} (\vec k+ \vec G)^2/2m + J\bm S_{\vec G'}\cdot \vec \sigma e^{i\vec G'\cdot \vec r} )u_{\vec k,\vec G}e^{i\vec G\cdot \vec r}\\
    H^{\text{R}}_{\vec k} u_{\vec k} &= \sum_{\vec G,\vec G',w} (\hbar \chi_w v_f \delta_{\vec G,\vec G'}  (\vec k + \vec G)    + J \bm S_{\vec G'} e^{i\vec G'\cdot \vec r}) \cdot \vec \sigma \,\, u_{\vec k,\vec G} e^{i\vec G\cdot \vec r}
\end{align}
\end{subequations}
which may be written as $[\mathsf{H}^{\text{NR}/\text{R}}_{\vec k}]_{\vec G,\vec G'} u_{\vec k,\vec G'} = E_{\vec k} u_{\vec k, \vec G}$ with 
\begin{subequations}
    \begin{align}
    [\mathsf{H}_{\vec k}^{\text{NR}}]_{\vec G,\vec G'} &= \delta_{\vec G,\vec G'}\frac{\hbar^2(\vec k + \vec G)^2}{2m}   + J \vec \sum_{\vec G''} \bm S_{\vec G''}\delta_{ \vec G'+ \vec G'' , \vec G}  \cdot \vec \sigma \\
        [\mathsf{H}_{\vec k}^{\text{R}}]_{\vec G,\vec G'} &= \sum_w (\hbar \chi_w v_f \delta_{\vec G,\vec G'} (\vec k + \vec G) + J \vec \sum_{\vec G''} \bm S_{\vec G''}\delta_{ \vec G'+ \vec G'' , \vec G} ) \cdot \vec \sigma.
    \end{align}
\end{subequations}
We solve this in a plane wave expansion and impose a cutoff $|\vec G| < G_{max}$. For convergence, we ensure $\hbar^2 G_{max}^2/2m\gg J$ and $\hbar v_f G_{max} \gg J$, respectively.\par 
The $\mathcal{P}$ ($\mathcal{T}$) operation connects Weyl points of the opposite (same) chirality~\cite{chan16}, so a single pair of Weyl points is allowed only in materials with broken ${\mathcal T}$-symmetry, where Faraday/Kerr rotations of light are also expected~\cite{trivedi15}. Both the GME and Faraday rotation refer to the rotation of the light polarization through the medium, though their origins are quite different. \par 
We augment the single pair of Weyl points with a second pair related to the first by $\mathcal{T}$. The Hall conductivities from the two pairs then cancel out, but their contributions to GME remain additive as the effect is even under $\mathcal{T}$. Although the coupling to the magnetic texture will eventually break $\mathcal{T}$, there will be no accompanying Faraday rotation as long as the coupling does not result in finite Hall conductivity. Though a minimum of two pairs of Weyl points are required to preserve $\mathcal{T}$, the GME calculation itself can be performed for a single pair and the final result multiplied by the total number of Weyl pairs.

\subsection{Gyrotropic magnetic effect}
We recall the formula for the $\alpha^{\text{GME}}$ tensor:
\begin{equation}\label{eq:alphaGMEappendix}
    \alpha^{\text{GME}}_{ij} = \frac{i\omega \tau e}{1-i\omega \tau} \int [d\vec k] (\partial f/\partial \varepsilon_{\vec k}) v_{\vec k,i} m_{\vec k,j}
\end{equation}
where we drop the band index for convenience. The magnetic moment is 
\begin{subequations}
    \begin{align}
    \vec m_{\vec k} &= \vec m_{\vec k}^{\text{spin}}+ \vec m_{\vec k}^{\text{orb}} \\
    &=  -\frac{eg_s}{2m_e}\bm s_{\vec k} + \frac{e}{2\hbar} \text{Im}\langle \bm \partial_{\vec k} u_{\vec k} | \times (H_{\vec k} - \varepsilon_{\vec k})|\bm \partial_{\vec k} u_{\vec k}\rangle
\end{align}
\end{subequations}
where $\bm s_{\vec k} = (\hbar /2) \langle u_{\bf k} | \vec\sigma |u_{\bf k} \rangle$ and $f(\varepsilon_{\vec k})$ is the occupation factor. At low temperatures, $\partial f/\partial \varepsilon_{\vec k} \approx - \delta(k-k_F)/\hbar |v_{\vec k}|$ and Eq. \eqref{eq:alphaGMEappendix} can be expressed as a surface integral. However, we find it useful to keep it as a volume integral for an easier numerical integration. Noting that $(\partial f/\partial \varepsilon_{\vec k}) v_{\vec k,i} = \hbar^{-1}(\partial f/\partial k_i)$ and integrating by parts, we have
\begin{equation}\label{eq:GMEtensorintbyparts}
    \agme_{ij} = -\frac{i\omega \tau e}{1-i\omega\tau}\frac{1}{\hbar} \int [d\vec k] f(\varepsilon_{\vec k}) \partial_{i} m_{\vec k,j}
\end{equation}
where $\partial_i \equiv \partial/\partial k_i$. Next, we put the magnetic moments into convenient forms.  It is useful to define the quantity
\begin{subequations}
    \begin{align}
        T_{\alpha\beta} &\equiv \langle \partial_\alpha u_{n,\vec k} | (H_{\vec k}-\epsilon_n)|\partial_\beta u_{n,\vec k}\rangle = \sum_m \braket{\partial_\alpha u_{n,\vec k}}{u_m} \braket{u_{m,\vec k}}{\partial_\beta u_{n,\vec k}}(\varepsilon_m-\epsilon_n)
    \end{align}
\end{subequations}
Using 
$
    \braket{u_{m,\vec k}}{\partial_\alpha u_{n,\vec k}} =\begin{cases}\frac{\braket{u_{m,\vec k}}{\partial_\alpha H_{\vec k}|u_{n,\vec k}}}{(\epsilon_n-\epsilon_m)} & n\neq m\\
    0 & n=m
    \end{cases}
$
we find
$
        T_{\alpha\beta} = \sum_{m\neq n} \frac{\braket{u_{n,\vec k}}{\partial_\alpha H_{\vec k}|u_{m,\vec k}} \braket{u_{m,\vec k}}{\partial_\beta H_{\vec k}|u_{n,\vec k}}}{(\epsilon_m-\epsilon_n)}.
$
Next, we define
\begin{subequations}
\begin{align}
        S_{\alpha\beta} &\equiv T_{\alpha\beta}-T_{\beta\alpha}= 2\sum_{m\neq n}\frac{\Im[\braket{u_{n,\vec k}}{\partial_\alpha H_{\vec k}|u_{m,\vec k}} \braket{u_{m,\vec k}}{\partial_\beta H_{\vec k}|u_{n,\vec k}}]}{(\epsilon_m-\epsilon_n)}.
\end{align}
\end{subequations}
For the non-relativistic case, 
\begin{equation}
S_{\alpha\beta}^{\text{NR}} = 2(\hbar^2/m)^2 \sum_{m\neq n} \frac{\text{Im}[\braket{u_{n,\vec k}}{G_\alpha|u_{m,\vec k}}\braket{u_{m,\vec k}}{G_\beta|u_{n,\vec k}}]}{(\varepsilon_m-\varepsilon_n)} . 
\end{equation}
The action of $G_\alpha$ on the Bloch state $u_{n,\vec k}(\vec r) = \sum_{\vec G} c_{n,\vec k,\vec G} e^{i\vec G\cdot \vec r}$ is understood as $\sum_{\vec G} c_{n,\vec k,\vec G} \vec G e^{i\vec G\cdot \vec r}=-i\nabla_{\vec r}u_{n,\vec k}(\vec r)$. 

For the relativistic case, 
\begin{equation}        S_{\alpha\beta}^{\text{R}}= 2(\hbar v_f)^2\sum_{m\neq n}\frac{\Im[\braket{u_{n,\vec k}|\sigma^\alpha}{u_{m,\vec k}} \braket{u_{m,\vec k}|\sigma^\beta}{u_{n,\vec k}}]}{(\epsilon_m-\epsilon_n)}
\end{equation}
The orbital magnetic moment is 
\begin{equation}
    m_{{\vec k},\alpha}^{\text{orb.}} =  \frac{e}{4\hbar}  \epsilon_{\alpha\beta\gamma} S_{\beta\gamma}^{\text{NR/R}},
\end{equation}
the spin magnetic moment in both cases is
\begin{equation}
    m_{\vec k,\alpha}^{\text{spin}} = -\frac{e\hbar g_s}{4m_e} \langle u_{\vec k} | \sigma^\alpha |u_{\vec k}\rangle.
\end{equation}
We evaluate $\partial_{i}m_{\vec k,j}$ numerically by finite difference, checking for convergence. We could also compute $\agme$ using Eq.~\eqref{eq:alphaGMEappendix}, in which case it is useful to note $v_{\vec k,\alpha}^{\text{NR}} = \frac{\hbar}{m}  \braket{u_{\vec k}}{ k_\alpha + G_\alpha|u_{\vec k}}$ and $v_{\vec k,\alpha}^{\text{R}} = \chi_w v_f \hbar \langle u_{\vec k} |\sigma^\alpha|u_{\vec k}\rangle$. The magnetic moments are calculated from using the complete set of Bloch eigenstates at each $\vec k$, obtained by diagonalizing the Hamiltonian in the plane wave basis with a momentum cutoff $|\vec G|<G_{max}$ ($\hbar^2G_{max}^2/2m\gg J$ and $\hbar v_f G_{max}\gg J$) to ensure convergence. A useful benchmark for the Weyl case is $J=0$, for which $m^{\text{orb.}}_{\vec k,\alpha} = -e\chi_wv_f k_\alpha/2k^2$ and $\alpha^{\text{GME,orb.}}_{ij} = -\chi_w\frac13\frac{e^2}{h^2}\varepsilon_f\delta_{ij}$ implying a rotatory power $\rho^{\text{orb.}} = \chi_w\frac{e^2}{3h^2c^2\varepsilon_0}\varepsilon_f$. 
\par 

In general, care must be taken when integrating by parts to produce Eq.~\eqref{eq:GMEtensorintbyparts}. Suppose for instance $f(k) = \Theta(a-k)$ defines a spherical Fermi surface of radius $a$. Suppose also that $\vec m_{\vec k} = c\hat{\vec k}/k^b$, for some power $b$ and constant $c$ in $d$ dimensions, and consider the integral
\begin{equation}\label{eq:Iij}
    I_{ij} \equiv \int d^d\vec k \, \partial_i f\, m_{\vec k,j} = -\frac{c}{d}\Omega_{d-1} \delta_{ij} a^{d-1-b}
\end{equation}
where $\Omega_d$ is the volume of the unit $d$-sphere. Na\"ively integrating by parts, we obtain 
\begin{equation}\label{eq:Iijstar}
    I_{ij}^* \equiv - \int d^d\vec k \,f\, \partial_i m_{\vec k,j} = -\frac{c}{d}\Omega_{d-1} \delta_{ij} a^{d-1-b} \qquad \text{if } b<d
\end{equation}
which diverges if $b\geq d$. One workaround is to split $I_{ij}$ into two integrals, the first over an $\epsilon$-ball centered at the origin and the second over its complement. The first can be treated exactly using Eq.~\eqref{eq:Iij} while the second can be treated exactly using integration by parts, keeping the boundary term. This procedure agrees with Eq.~\eqref{eq:Iijstar} if $b<d$, in which case $\epsilon$ can be taken to $0$. In all integrals we study, the analog of $I^*_{ij}$ converges (i.e. $b<d$), so we do not worry about this subtlety. Nevertheless, it is sometimes useful to remove an $\epsilon$-ball from the origin to aid numerical stability and Eq.~\eqref{eq:Iij} quantifies the corresponding error. \par 
Moreover, when the support of the distribution function $f$ is unbounded (as happens for the helical spin spiral with a hole-like Fermi surface), it is convenient to replace $f$ with $-(1-f)$, which manifestly preserves Eq.~\eqref{eq:Iij} and simplifies numerical calculation of Eq.~\eqref{eq:Iijstar}. These considerations are mainly useful for the Weyl semimetal setting. \par 

We list here some other numerical details. For the non-relativistic case, in the regime of small $J/E_K$, the band bottom varies as $\varepsilon_{\text{min}} \approx -J$ due to the Hunds coupling. Therefore we fix $E_F - \varepsilon_{\text{min}}$ (the natural notion of Fermi energy) rather than $E_F$ itself. For the Weyl semimetal in a spin spiral, we excise a small $\epsilon$-ball around $k_\perp = 0$ and replace $f$ with $-(1-f)$ for the hole-like Fermi surface case to improve numerical convergence. For the same case, we found numerical convergence to be slow with the number $N_k^3$ of momentum grid points so finite-size scaling was employed. For this, linear fits were obtained to the data as a function of $1/N_k^3$ extrapolated to $N_k=\infty$. In other cases, finite-size scaling was not necessary. Results were checked for convergence with respect to $G_{max}$ and $N_k$. Temperature was set to $k_B T = 1$ meV. 
\par 

The GME occurs in materials with broken $\mathcal{P}$ while Faraday rotation occurs when the broken $\mathcal{T}$ results in finite Hall conductivity. Here we focus on the GME, augmenting the single pair of Weyl points with a second pair related to the first by $\mathcal{T}$. The Hall conductivities from the two pairs then cancel out, but their contributions to GME remain additive as the effect is even under $\mathcal{T}$. Although the coupling to the magnetic texture will eventually break $\mathcal{T}$, there will be no accompanying Faraday rotation as long as the coupling does not result in finite Hall conductivity. Though a minimum of two pairs of Weyl points are required to preserve $\mathcal{T}$, the GME calculation itself can be performed for a single pair and the final result multiplied by the total number of Weyl pairs. 

\section{Details of multi-$q$ spin texture}
\label{sec:multiq}
In this section we include the exact form of the multi-$q$ spin spiral used in the main text. For concreteness, we adopted the face-centered cubic (FCC) multi-$q$ spiral texture of Ref. \cite{Park2011May}, among the most energetically favorable, defined as
\begin{equation}
    \begin{aligned}
        \bm S_{\vec r} &= \bm S_{\vec 0} + \sum_{j} \left[\bm S_{\vec G_j} e^{i\vec G_j \cdot \vec r} +  (\bm S_{\vec G_j})^* e^{-i\vec G_j \cdot \vec r}\right]\\
        \bm S_{\vec G_1} &= \frac{1}{4} e^{i\theta_1 }\left(\frac{\hat x + \hat y -2\hat z}{\sqrt{6}} +i\chi_s \frac{\hat y - \hat x}{\sqrt{2}}\right)\\
        \bm S_{\vec G_2} &= \frac{1}{4} e^{i\theta_2}\left(\frac{\hat x + \hat y +2\hat z}{\sqrt{6}} +i\chi_s \frac{\hat y - \hat x}{\sqrt{2}}\right)\\
        \bm S_{\vec G_3}&= \frac{1}{4} e^{i\theta_3}\left(\frac{\hat x + 2\hat y +\hat z}{\sqrt{6}} +i\chi_s \frac{\hat x - \hat z}{\sqrt{2}}\right)\\
        \bm S_{\vec G_4} &= \frac{1}{4} e^{i\theta_4}\left(\frac{2\hat x + \hat y +\hat z}{\sqrt{6}} +i\chi_s \frac{\hat z - \hat y}{\sqrt{2}}\right)
    \end{aligned}
\end{equation}
where $\hat{\vec G}_1 = (\hat x+\hat y+\hat z)/\sqrt{3}, \hat{\vec G}_2 = (-\hat x -\hat y+\hat z)/\sqrt{3}, \hat{\vec G}_3 = (-\hat x + \hat y-\hat z)/\sqrt{3}, \hat{\vec G}_4 = (\hat x-\hat y-\hat z)/\sqrt{3}$, the $\theta_i$ are free parameters, and $\chi_s$ is again the spin chirality. We set $\theta_i = 0$.\par 

\section{Bandstructures}
\label{sec:bandstructures}

\begin{figure}
    \centering
    \includegraphics[width=\linewidth]{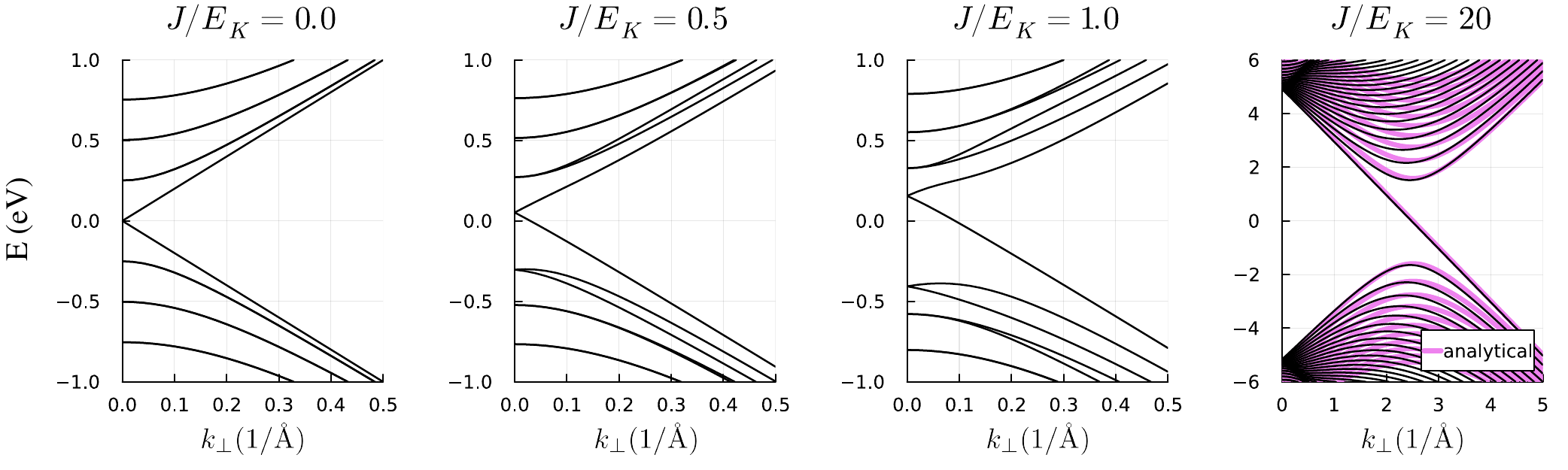}
    \caption{\textbf{Spin spiral bandstructure (relativistic).} Bandstructure of Weyl electrons coupled to a spin spiral $\bm{S}_{\vec r} = (\cos 2\pi z/\lambda,\chi_s \sin 2\pi z/\lambda,0)$ for varying $J/E_K$, where $J$ is the Hund's coupling and $E_K = hv_F/\lambda$ is the characteristic electronic kinetic energy. Here $k_\perp = |(k_x,k_y)|$ is the transverse momentum, while bands are quite flat in the periodic momentum $k_z$ (not shown) at any $J\gtrsim E_K$. On right, analytical approximate solutions derived in the large $J$ limit (Eq.~\eqref{localized-solution}). The linearly dispersing midgap modes are Gaussian-localized along $z$. Parameters: $\chi_s=\chi_W = 1$, $\hbar v_F = 0.2$ eV nm (or $v_F \approx 10^{-3}c$), $\lambda = 5$ nm.}
    \label{fig:weylhelical}
\end{figure}

\begin{figure}
    \centering
    \includegraphics[width=\linewidth]{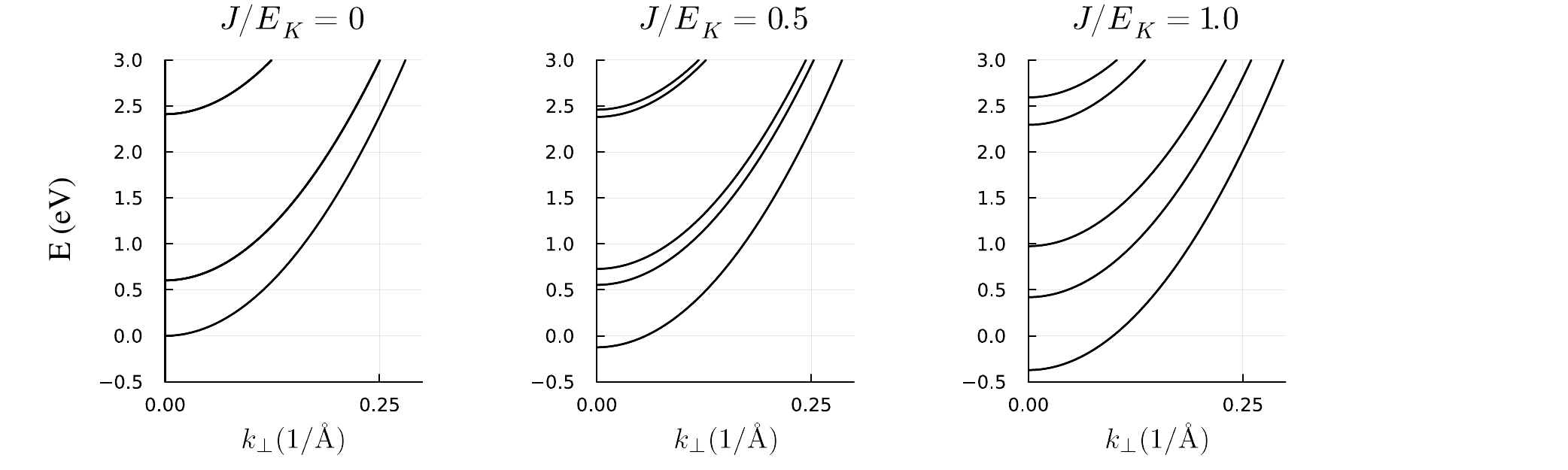}
    \caption{\textbf{Spin spiral bandstructure (non-relativistic).} Bandstructure of electrons coupled to a spin spiral $\bm{S}_{\vec r} = (\cos 2\pi z/\lambda,\chi_s \sin 2\pi z/\lambda,0)$ for varying $J/E_K$, where $J$ is the Hund's coupling and $E_K = h^2/2m\lambda^2$ is the characteristic electronic kinetic energy. Here $k_\perp = |(k_x,k_y)|$ is the transverse momentum. Parameters: $\chi_s = \chi_w = 1$, $m = 0.1 m_e$, $\lambda = 5$ nm. }
    \label{fig:numericshelical}
\end{figure}

\begin{figure}
    \centering
    \includegraphics[width=\linewidth]{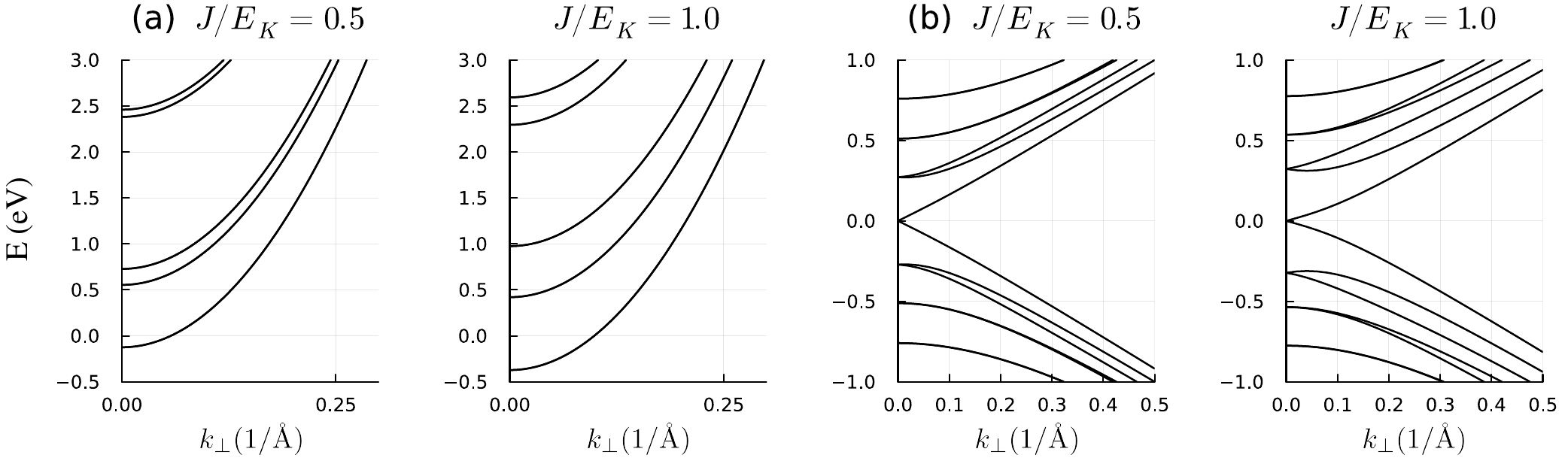}
    \caption{\textbf{Spin cycloid bandstructure.} Bandstructure of (a) non-relativistic and (b) Weyl electrons coupled to a spin cycloid $\bm{S}_{\vec r} = (\cos 2\pi z/\lambda,0,\chi_s \sin 2\pi z/\lambda)$ for varying $J/E_K$, where $J$ is the Hund's coupling and $k_\perp = |(k_x,k_y)|$ is the transverse momentum. Parameters: $\chi_s = \chi_w = 1$, $m = 0.1 m_e$, $\hbar v_F = 0.2$ eV nm, $\lambda = 5$ nm. }
    \label{fig:cycloid}
\end{figure}

\begin{figure}
    \centering
\includegraphics[width=\linewidth]{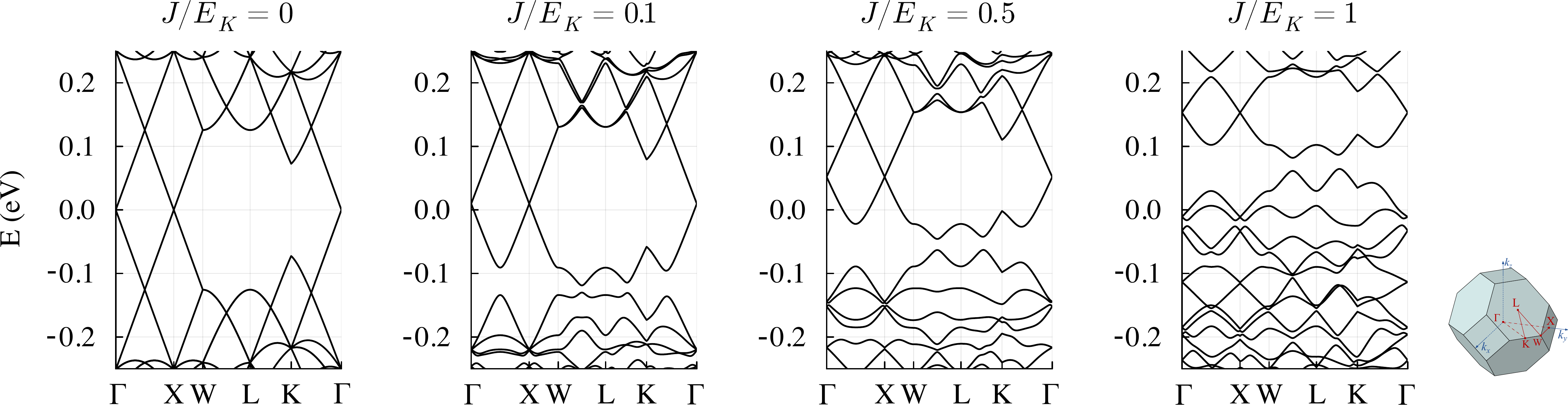}    \caption{\textbf{Multi-$q$ bandstructure (relativistic).} Bandstructure of Weyl electrons coupled to a FCC multi-$q$ spin texture (of wavelength $\lambda$) for varying $J/E_K$, where $J$ is the Hund's coupling and $E_K = hv_F/\lambda$ is the characteristic electronic kinetic energy. Parameters: $\chi_s = \chi_w = 1$, $\hbar v_F = 0.2$ eV nm (or $v_F \approx 10^{-3}c$), $\lambda = 5$ nm. }
    \label{fig:numerics_bandstructure}
\end{figure}

\begin{figure}
    \centering
\includegraphics[width=\linewidth]{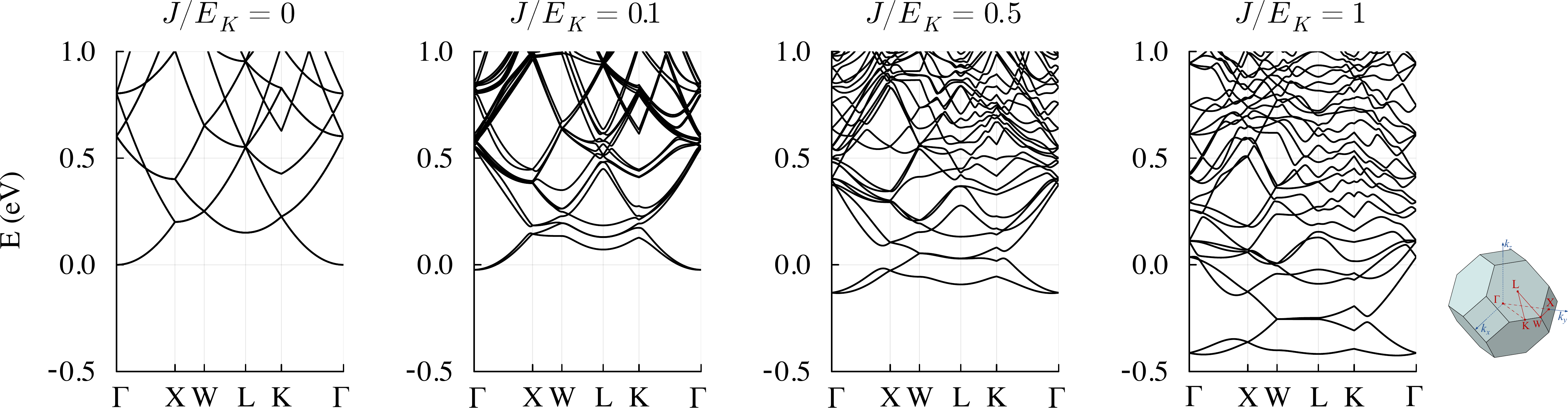}
    \caption{\textbf{Multi-$q$ bandstructure (non-relativistic).} Bandstructure of electrons coupled to a FCC multi-$q$ spin texture (of wavelength $\lambda$) for varying $J/E_K$, where $J$ is the Hund's coupling and $E_K = h^2/2m\lambda^2$ is the characteristic electronic kinetic energy. Here $k_\perp = |(k_x,k_y)|$ is the transverse momentum. Parameters: $\chi_s = \chi_w = 1$, $m = 0.1 m_e$, $\lambda = 5$ nm. }
    \label{fig:numericsmultiq_bandstructure}
\end{figure}

We plot some representative bandstructures for the spin textures considered in this work. We refer to Appendix \ref{app:alphanumerical} for numerical details. A summary of the figures is as follows:
\begin{itemize}
    \item Figures~\ref{fig:weylhelical} and \ref{fig:numericshelical}: spin spiral texture
    \item Figure~\ref{fig:cycloid}: spin cycloid texture
    \item Figures~\ref{fig:numerics_bandstructure} and \ref{fig:numericsmultiq_bandstructure}: FCC multi-$q$ texture
\end{itemize}

\section{Analytical approach to the WSM coupled to the spin spiral} 
\label{eq:WSMspiral}
\subsection{Zero modes}

The Weyl semimetal coupled to the spin spiral admits an approximate analytical solution. Without loss of generality, we suppose the spiral propagates along $\hat{\vec G} = \hat z$ and write this model as
\begin{equation}
    H = \sum_w \chi_w v_F(\vec p\cdot \bm \sigma)_w + J \sum_{\vec r} \bm S_{\vec r}\cdot \bm \sigma ,\quad \bm S_{\vec r} = (\cos Gz,\chi_s \sin Gz,0)
\end{equation}
with $J>0$. We assume two Weyl nodes $w=\pm$ with $\chi_{\pm} = \pm 1$. We treat transverse momentum as a quantum number $ {\bf k}_\perp = (k_x , k_y)$ and write the Hamiltonian, keeping the $\sum_w$ implicit, as
\begin{equation}
    H  = -i\chi_w\hbar v_F\sigma^z \partial_z + M_x \sigma^x + M_y \sigma^y
    \label{H-kxky}
\end{equation}
where ${\bm M} =(M_x,M_y) =\chi_w\hbar v_F {\bf k}_\perp +J (S_{\vec r}^x,S_{\vec r}^y)$ is a Dirac mass. We make progress by linearizing around the minima of
\begin{equation}
    M^2 = (\hbar v_F k_\perp)^2 + J^2 +2\chi_w\hbar v_Fk_\perp J\cos\left(\theta_\perp +\chi_s Gz\right)
\end{equation}
where $k_x+ik_y = k_\perp e^{i\theta_\perp}$. The minima occur at $Gz_n = 2\pi \left( n+\frac{\delta_{\chi_w,1}}{2}\right) -\theta_\perp \chi_s$. Near each position $z_n$, localized modes may be described to first order in $z-z_n$ by the Hamiltonian
\begin{align}
    H  \approx  -i \chi_w \hbar v_F\sigma^z \partial_z   + (\chi_w e^{i\theta_\perp} \{ \hbar v_f k_\perp - J [ 1 + i \chi_s  G(z-z_n ) ] \} \sigma^- + \text{H.c.}).
    \label{H-linearized}
\end{align}
The phase factors $e^{\pm i \theta_\perp}$ can be absorbed through rotation $\sigma^\pm \rightarrow \sigma^\pm e^{\pm i \theta_\perp}$:
\begin{align}
H \approx  \chi_w \Bigl[ -i \hbar v_F\sigma^z \partial_z + ( \hbar v_F k_\perp - J ) \sigma^x  - \chi_s J G (z-z_n ) \sigma^y  \Bigr] . 
    \label{H-linearized2}
\end{align}
Next, we assume that the eigenstates of $H$ are simultaneous eigenstates of $\sigma^x$ with eigenvalue $s$:
\begin{align}\label{eq:linearizedH}
H \approx  i \chi_w \sigma^z \Bigl[ -\hbar v_F \partial_z  + s  \chi_s J G(z-z_n) \Bigr] + s \chi_w ( \hbar v_F k_\perp - J ) . 
\end{align}
For $s=-\chi_s$, we find Jackiw-Rebbi-type~\cite{jackiw-rebbi} solutions 
\begin{align} \phi_n (z)  & = (\pi l_1^2 )^{-1/4} e^{- (z-z_n )^2 / 2l_1^2 }\hat \chi , \label{localized-solution} \nn
\varepsilon_{k_\perp} & = -\chi_w\chi_s ( \hbar v_F k_\perp  -J ) .
\end{align}
where $l_1^2 =  \hbar v_F / GJ$ and $\hat\chi$ is the appropriate normalized spinor. The higher bands are obtained in the same approximation by squaring Eq.~\eqref{H-linearized2} to get
\begin{equation}
    H^2 = -(\hbar v_F)^2 \partial_z^2 + (\hbar v_Fk_\perp - J)^2 + (JG)^2(z-z_n)^2 +\chi_s JG\hbar v_F \sigma^x
\end{equation}
and solving the resulting harmonic oscillator problem. Letting $\sigma^x = s = \pm 1$, the (squared) energy levels are 
\begin{equation}
    \epsilon_{n,s}^2 = 2\hbar v_F JG (n+(1+s\chi_s)/2) + (\hbar v_F k_\perp - J)^2.
\end{equation}
This solution reduces to the previous case when the first term vanishes. The low-lying spectrum is decently borne out by the exact numerical solution in the large $J/E_K$ limit, as we show in Fig.~\ref{fig:weylhelical}. It is interesting to note that the $z$-positions of these modes are ``locked" to $\theta_\perp$. \par 
Despite these attractive features, the solutions we obtained in \eqref{localized-solution} do not obey a certain ``gauge invariance" property and should be treated with caution. It is perfectly conceivable that a unitary transformation $U(\alpha ) = e^{i \alpha \chi_w\chi_sGz\sigma^z /2}$ with arbitrary phase $\alpha$ be implemented on the Hamiltonian $H \rightarrow U(\alpha) H U^\dag (\alpha)$ before the linearization. Taking $\alpha = 1$, $U\equiv U(1)$, gives
\begin{align}
\tilde H & = U H U^\dag =  -i\chi_w\hbar v_F\sigma^z\partial_z - J \sigma^x + \chi_w \hbar  v_F k_\perp  \left[ e^{i(\chi_sGz-\theta_\perp)}\sigma^++\text{H.c.} \right]  - \chi_s\chi_w\hbar v_FG/2. 
\label{H-reduced} 
\end{align}
However, proceeding as before (expanding to linear order around the minima of the Dirac mass), we find instead the localized modes and dispersion
\begin{align} 
\phi_n (z) & = ( \pi l_2^2 )^{-1/4} e^{-(z-z_n)^2 /2 l_2^2 } \hat \chi , \nn 
\varepsilon_{k_\perp} & = -\chi_w\chi_s (\hbar v_F k_\perp -J) - \chi_w \chi_s \hbar v_FG/2
\label{localized-solution2}
\end{align}
where $l_2^2 = 1/G k_\perp$, in contrast with \eqref{localized-solution}. The lack of equivalence between the two solutions obtained is a consequence of the noncommutativity of the gauge transformation and linearization step. Interestingly, we find that the numerically obtained eigenfunctions are well fit (over several-orders-of-magnitude ranges for $k_\perp$ and $J$) by 
\begin{equation}
    \phi_n(z) = ( \pi l^2 )^{-1/4} e^{-(z-z_n)^2 /2 l^2 } \hat \chi
    \label{localized-solution3}
\end{equation}
where $l=c(l_1l_2)^{1/2}$ is proportional to the \textit{geometric mean} of $l_1$ and $l_2$ (and $c\approx 0.5$). The numerically obtained energy is well-fit by both Eq.~\eqref{localized-solution} and Eq.~\eqref{localized-solution2} up to corrections which are small compared to $\hbar v_F k_\perp$ and $J$. \par

\subsection{Magnetic moments and large$-J/E_K$ rotatory power}

For the WSM coupled to the spin spiral, excellent numerical fits were found to the following forms of spin and orbital magnetic moment in the large $J/E_K$ regime and $\hbar v_F k_\perp \lesssim J$ for the linearly dispersing band discussed in the previous section. They are: 
\begin{equation}\label{eq:largeJmom}
    \begin{aligned}
    \vec m_{\vec k}^{\text{spin}} &\approx \frac{-eg_s}{2m_e} \frac{\hbar}{2} \chi_s \hat{\vec k}_\perp\\
    \vec m_{\vec k}^{\text{orb}} &\approx \frac14 \chi_w e v_f \frac{\hat{\vec{k}}_{\perp}}{(J/\hbar v_F)^{1/3}k_\perp^{2/3}}
    \end{aligned}
\end{equation}
Moreover, the dispersion $\varepsilon_{\vec k} = -\chi_w \chi_s(\hbar v_F k_\perp - J)$ fits well with the numerical solution up to subleading terms which are safe to ignore in this regime. We will calculate the rotatory power assuming this dispersion and Eq.~\eqref{eq:largeJmom} (i.e. ignoring subleading corrections). Noting $\partial_i f(\varepsilon_{\vec k})= \partial_i \Theta[\chi_s\chi_w(k - k_f)] = \hat{k}_i \, \chi_w \chi_s\delta(k-k_f)$ at zero temperature, we have  
\begin{equation}
    \begin{aligned}
\alpha_{ij}^{\text{GME,R,spin}} &= -\frac{e}{\hbar}\sum_{w} \int [d\vec k] \partial_i f(\varepsilon_{\vec k}) m_{\vec k,j}^{\text{spin}} \\
        &= \frac{eG}{2 (2\pi)^3} \frac{eg_s}{2m_e}  \sum_{w} \chi_w \int \, k_\perp \, dk_\perp \, d\theta\, \delta(k_\perp -k_f) \hat k_{\perp,i} \hat k_{\perp,j} \\
        &= \frac{-\chi_se^2Gg_s\varepsilon_f}{8\pi m_e h v_f} \hat\delta_{ij}
    \end{aligned}
\end{equation}
where $\hat\delta_{ij} = \text{diag}(1,1,0)$. The rotatory power is
\begin{equation}
    \begin{aligned}       \rho_{\|}^{\text{R,spin}} &= \chi_s \alpha_{\text{FS}}  \frac{g_sE_FG}{4\pi  m_e  v_F c} 
    \end{aligned}
\end{equation}
and $\rho_{\perp}^{\text{R,spin}} = \rho_{\|}^{\text{R,spin}} /2$. Meanwhile, 
\begin{equation}
    \begin{aligned}
        \alpha_{ij}^{\text{GME,R,orb}} &= -\frac{e}{\hbar} \sum_{\chi_w} \int [d\vec k] \partial_i f(\varepsilon_{\vec k})m_{\vec k,j}^{\text{orb}} \\
        &= -\frac{e^2\chi_sv_fG}{4\hbar (J/\hbar v_f)^{1/3}(2\pi)^3} \sum_{\chi_w} \int k_\perp^{1/3} \,dk_\perp\,d\theta\, \delta(k-k_f) \hat k_{\perp,i} \hat k_{\perp,j} \\
        &= \frac{-\chi_s}{4}\frac{ e^2}{h^2}  (\hbar v_f G) \hat\delta_{ij}   + O(E_F^2/J^{2})
    \end{aligned}
\end{equation}
and the rotatory power is 
\begin{equation}
    \begin{aligned}
        \rho_{\|}^{\text{R,orb}} &= \chi_s \alpha_{\text{FS}} \frac{v_F G}{4\pi c} 
    \end{aligned}
\end{equation}
assuming $E_F \ll J$, with $\rho_{\perp}^{\text{R,orb}} = \rho_{\|}^{\text{R,orb}} /2$.

\section{Materials \& estimates}

In this section we identify a few material candidates for unambiguous observation of the GME induced by chiral magnetic textures. In order to unambiguously attribute the GME to the chiral magnetic texture, it is desirable to find candidates which are centrosymmetric in their paramagnetic phases and only become globally non-centrosymmetric upon magnetic ordering. Examples of such materials are centrosymmetric skyrmion materials or centrosymmetric helimagnets. The origin of chiral spin textures in centrosymmetric materials is a frustrated longer-range spin-spin interactions such the RKKY interaction induced by itinerant electrons, rather than the Dzyaloshinskii–Moriya interaction (DMI) common to non-centrosymmetric magnets. Below we list a few candidates and their salient features:
\begin{itemize}
    \item Gd$_2$PdSi$_3$~\cite{kurumaji2019skyrmion,hirschberger2020high}: frustrated centrosymmetric triangular-lattice magnet with a field-induced Bloch-type skrymion lattice phase, confirmed by a giant topological Hall response and corroborated by resonant x-ray scattering, with a period $\lambda \sim 2.4$ nm; also, a zero field helical spiral or triple-q state. 
    \item Gd$_3$Ru$_4$Al$_{12}$ \cite{hirschberger2019skyrmion}: frustrated centrosymmetric breathing kagome lattice magnet with field-induced Bloch-type skyrmion lattice and zero field helical spiral, confirmed by topological Hall response, resonant x-ray scattering, small angle neutron scattering, and Lorentz TEM, with a period $\lambda \sim 2.8$ nm. 
    \item GdRu$_2$Si$_2$~\cite{khanh2020nanometric}: centrosymmetric tetragonal magnet with field-induced double-q square skyrmion lattice and zero field helical spiral, as confirmed by resonant x-ray scattering and Lorentz TEM, with period $\lambda \sim 1.9$ nm. 
    \item MnAu$_2$~\cite{herpin1961etude,masuda2024room}: centrosymmetric tetragonal magnet with zero field helical spiral of period $\lambda \sim 3.1$ nm. 
    \item MnP~\cite{moon1982neutron,PhysRevB.93.100405}: centrosymmetric orthorombic magnet with zero field helical spiral
\end{itemize}

\begin{figure}[t]
  \centering
  \begin{minipage}[t]{0.6\linewidth}\vspace{0pt}%
    \small
    \setlength{\tabcolsep}{6pt}
    \begin{tabular}{|c|c|c|}
      \hline
      \textbf{Helical} & \textbf{SkL } & \textbf{Centrosymmetric} \\
      \hline &&\\
      $\rho_\|^{\text{NR}}=0$ &
      $\rho_z^{\text{NR}}=\pm4.9\times10^{-1}$ &
      $\rho^{\text{NR}}=0$ \\
      $\rho_\perp^{\text{NR}}=\pm 2.0\times10^{-1}$ &
      $\rho_{\text{in-plane}}^{\text{NR}}=\pm 2.4\times10^{-1}$ &
      \\
      \hline
    \end{tabular}
  \end{minipage}
  \begin{minipage}[t]{0.3\linewidth}\vspace{-30pt}%
    \centering
    \includegraphics[width=\linewidth]{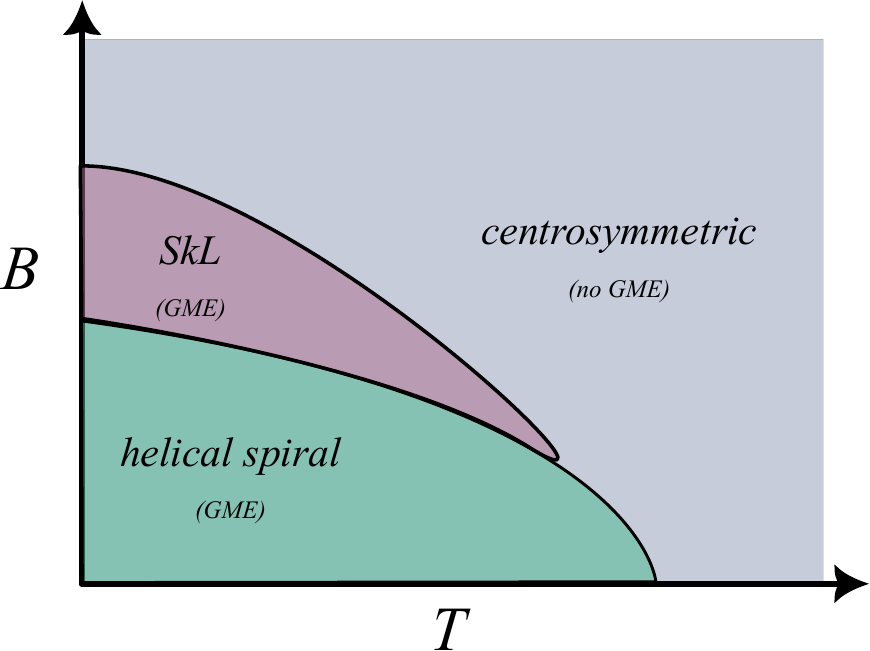}
  \end{minipage}
  \caption{{\textbf{Estimates for Gd$_2$PdSi$_3$. } (Left) Estimates for rotatory power in radians / mm in the phases of Gd$_2$PdSi$_3$ using parameters from Ref.~\cite{chen2025}. See text for details. (Right) Schematic of phase diagram typical of frustrated centrosymmetric magnets including Gd-based compounds, with a helical spiral and SkL phase at low temperature and field and centrosymmetric paramagnetic or field-aligned phases at larger fields and temperatures.}}
  \label{fig:rotatory_table_plus_graphic}
\end{figure}

As a representative example, we provide an estimate of the GME effect in Gd$_2$PdSi$_3$. Gd$_2$PdSi$_3$ hosts at least three phases under applied field~\cite{kurumaji2019skyrmion,hirschberger2020high,paddison}: a phase dubbed IC-1 at zero and small applied field, which may be either a helical spiral or a triple-q noncoplanar meron-antimeron texture, a skyrmion lattice (SkL) at intermediate field 0.5-1 T which hosts a giant topological Hall response, and centrosymmetric phases at higher fields and temperatures. See Fig.~\ref{fig:rotatory_table_plus_graphic} for a schematic phase diagram. We will focus on a helical spiral order at zero field and the SkL, leaving aside the putative meron-antimeron texture.  We remark that there is still some debate about the zero-field low-temperature phase of Gd$_2$PdSi$_3$, with the two main candidates being the helical spiral and triple-q meron-antimeron texture, and the GME is well suited to distinguishing these two states due to the differing symmetries of the GME tensor depending on the presence of $C_3$ symmetry. \par 
\par 

\textbf{Helical phase. } In the helical phase, only the perpendicular component of the rotatory power is nonvanishing:
\begin{equation}
    \rho_\|^{\text{NR}} = 0,\qquad \rho^{\text{NR}}_\perp = \chi_s \frac{\rho_0}{2} \frac{g_s J}{4m_e(v_F^{\text{NR}})^2}g^{\text{NR}}(\xi)  
\end{equation}
with $\chi_s=\pm 1, \rho_0 = \alpha_0/c^2\varepsilon_0,v_F^{\text{NR}} = \hbar k_F/m,\xi = 2k_F/G, \alpha_0=e^2J/h^2,$ and $G=2\pi/\lambda$.

\textbf{SkL phase. } For the SkL phase, we adopt the spin texture from Ref.~\cite{chen2025}: 
\begin{equation}
\begin{aligned}
    \bm S_{\vec r} &= S_{0} \hat z + \sum_{i=1,2,3}(\bm S_i e^{i\vec Q_i\cdot \vec r} + \text{c.c.}) \\
    \bm S_i &= s_z \hat z + i\chi_s s_\perp \hat{\vec Q}_i\times \hat z,
\end{aligned}
\end{equation}
with $\vec Q_1 = \frac{4\pi}{\sqrt{3}\lambda}(-\sqrt{3}/2,-1/2,0), \vec Q_2 = \frac{4\pi}{\sqrt{3}\lambda}(\sqrt{3}/2,-1/2,0),\vec Q_3 = \frac{4\pi}{\sqrt{3}\lambda}(0,1,0)$. We take $S_0 = 0, s_z=s_\perp = -1/\sqrt{2}$. The form factors $f_{\vec G , ij } = i \vec{\hat{G}}_i (\bm S_{\vec G} \times \bm S_{-\vec G})_j$ take the form
\begin{equation}
    \begin{aligned}
        f_{\vec Q_a,ij} &= \chi_s \hat{ Q}_{a,i}\hat{ Q}_{a,j}
    \end{aligned}
\end{equation}
for $a=1,2,3$. The GME tensor in the clean limit takes the form
\begin{equation}    \alpha_{ij}^{\text{GME,NR,spin}} = -3\chi_s \alpha_0 \frac{g_sJ}{4m_e(v_F^{\text{NR}})^2} g^{\text{NR}}(\xi) \hat\delta_{ij}
\end{equation}
where $\hat \delta_{ij} = \text{diag}(1,1,0)$. The rotatory power is
\begin{equation}
    \rho_{z}^{\text{NR}} = 3\chi_s \rho_0\frac{g_sJ}{4m_e(v_F^{\text{NR}})^2}g^{\text{NR}}(\xi),\qquad \rho_{\text{in-plane}}^{\text{NR}} = \rho_{z}^{\text{NR}}/2. 
\end{equation}

\textbf{Centrosymmetric phase. } Finally, in the centrosymmetric phases at high $B$ field and high temperatures, the GME response vanishes: 
\begin{equation} \rho^{\text{NR}} = 0. 
\end{equation}

First-principles calculation of the bandstructure of Gd$_2$PdSi$_3$ indicates an approximately 500 meV spin splitting of conduction bands due to the local magnetic moments~\cite{chen2025}, and hence we take the Hunds coupling to be $J=250$ meV. We also adopt a period $\lambda \sim 2.4$ nm and the estimates $m = 0.5m_e$ and $E_F = 50$ meV. We present the corresponding rotatory power estimates in Fig.~\ref{fig:rotatory_table_plus_graphic}. The rotatory powers are well within the observable range and quite enhanced relative to the values quoted in the main text for the nonrelativistic metallic chiral magnet due to the comparatively larger Hunds coupling $J$ and smaller period $\lambda$. 

In conclusion, frustrated centrosymmetric magnets are ideal candidates for a sharp demonstration of the GME. By global symmetry considerations, the GME vanishes identically in the centrosymmetric phases at high temperature and field and only becomes nonvanishing in the chiral magnet phases (see Fig.~\ref{fig:rotatory_table_plus_graphic} for a schematic phase diagram). Additionally, the short few-nanometer period typical of these materials and large Hunds coupling evidenced by topological Hall effects situates them ideally for the observation of sizable rotatory powers.

\bibliography{ref}